%%  Last update & correction:  PC 26 Jan 00
%
%
%%%%%%%%%%%%%%%%%%%%%%%%%%%%%%%%%%%%%%%%%%%%%%%%%%%%%%%%%%%%%%%%%%%%%%%%%%%%%%
%%%%%%%%%%%%%%%%%%%%%%%%%%%%%[ Labelling    ]%%%%%%%%%%%%%%%%%%%%%%%%%%%%%%%%%
%%%%%%%%%%%%%%%%%%%%%%%%%%%%%%%%%%%%%%%%%%%%%%%%%%%%%%%%%%%%%%%%%%%%%%%%%%%%%%
%
\expandafter \def \csname CHAPLABELintro\endcsname {1}
\expandafter \def \csname EQLABELduals\endcsname {1.1?}
\expandafter \def \csname CHAPLABELtoric\endcsname {2}
\expandafter \def \csname EQLABELchiphys\endcsname {2.1?}
\expandafter \def \csname EQLABELhnos\endcsname {2.2?}
\expandafter \def \csname EQLABELchiBat\endcsname {2.3?}
\expandafter \def \csname EQLABELdivisors\endcsname {2.4?}
\expandafter \def \csname EQLABELcrossterms\endcsname {2.5?}
\expandafter \def \csname CHAPLABELheterotic\endcsname {3}
\expandafter \def \csname EQLABELmanifolds\endcsname {3.1?}
\expandafter \def \csname TABLABELnablas\endcsname {3.1?}
\expandafter \def \csname TABLABELbases\endcsname {3.2?}
\expandafter \def \csname EQLABELXfibrations\endcsname {3.2?}
\expandafter \def \csname EQLABELinjections\endcsname {3.3?}
\expandafter \def \csname EQLABELprojections\endcsname {3.4?}
\expandafter \def \csname EQLABELmirrordual\endcsname {3.5?}
\expandafter \def \csname CHAPLABELX\endcsname {4}
\expandafter \def \csname EQLABELcharges\endcsname {4.1?}
\expandafter \def \csname FIGLABELfanfigs\endcsname {4.1?}
\expandafter \def \csname FIGLABELBfig\endcsname {4.2?}
\expandafter \def \csname FIGLABELKY\endcsname {4.3?}
\expandafter \def \csname FIGLABELtriangs\endcsname {4.4?}
\expandafter \def \csname CHAPLABELyuk\endcsname {5}
\expandafter \def \csname EQLABELineqs\endcsname {5.1?}
\expandafter \def \csname EQLABELbasis\endcsname {5.2?}
\expandafter \def \csname EQLABELmcone\endcsname {5.3?}
\expandafter \def \csname EQLABELfiber\endcsname {5.4?}
\expandafter \def \csname EQLABELparams\endcsname {5.5?}
\expandafter \def \csname EQLABELvolE\endcsname {5.6?}
\expandafter \def \csname CHAPLABELsuperpot\endcsname {6}
\expandafter \def \csname CHAPLABELappZ\endcsname {-2}
\expandafter \def \csname FIGLABELZfig\endcsname {-2.1?}
\expandafter \def \csname CHAPLABELappY\endcsname {-3}
%
%%%%%%%%%%%%%%%%%%%%%%%%%%%%%%%%%%%%%%%%%%%%%%%%%%%%%%%%%%%%%%%%%%%%%%%%%%%%%%
%%%%%%%%%%%%%%%%%%%%%%%%%%%%[ xenia.mac     ]%%%%%%%%%%%%%%%%%%%%%%%%%%%%%%%%%
%%%%%%%%%%%%%%%%%%%%%%%%%%%%%%%%%%%%%%%%%%%%%%%%%%%%%%%%%%%%%%%%%%%%%%%%%%%%%%
%
\magnification=1200
%%%%%%%%%%%%%%%%%%%%%%%%%%%%%%%%%%%%%%%%%%%%%%%%%%%%%%%%%%%%%%%%%%%%%%%%%%%%%%%
%Fonts%%%%%%%%%%%%%%%%%%%%%%%%%%%%%%%%%%%%%%%%%%%%%%%%%%%%%%%%%%%%%%%%%%%%%%%%%
%%%%%%%%%%%%%%%%%%%%%%%%%%%%%%%%%%%%%%%%%%%%%%%%%%%%%%%%%%%%%%%%%%%%%%%%%%%%%%%

\font\eightrm=cmr8 at 8pt
\font\fourteenrm=cmr12 at 14pt
\font\seventeenrm=cmr17 at 17pt
\font\twentyonerm=cmr17 at 21pt

\font\ss=cmss10

\font\csc=cmcsc10

\font\twelvecal=cmsy10 at 12pt

\font\twelvemath=cmmi12
\font\fourteenmath=cmmi12 at 14pt 

\font\fourteenbold=cmbx12 at 14pt
\font\seventeenbold=cmbx7 at 17pt

\font\fively=lasy5
\font\sevenly=lasy7
\font\tenly=lasy10

\textfont10=\tenly
\scriptfont10=\sevenly    
\scriptscriptfont10=\fively
%%%%%%%%%%%%%%%%%%%%%%%%%%%%%%%%%%%%%%%%%%%%%%%%%%%%%%%%%%%%%%%%%%%%%%%%%%%%%%%
%Formatting%%%%%%%%%%%%%%%%%%%%%%%%%%%%%%%%%%%%%%%%%%%%%%%%%%%%%%%%%%%%%%%%%%%%
%%%%%%%%%%%%%%%%%%%%%%%%%%%%%%%%%%%%%%%%%%%%%%%%%%%%%%%%%%%%%%%%%%%%%%%%%%%%%%%
%\magnification=1200
\parskip=10pt
\parindent=20pt
\def\today{\ifcase\month\or January\or February\or March\or April\or May\or June
       \or July\or August\or September\or October\or November\or December\fi
       \space\number\day, \number\year}

\def\title#1{\footline={\ifnum\pageno<2\hfil
       \else\hss\tenrm\folio\hss\fi}\vskip1truein\centerline{{#1}   
       \footnote{\raise1ex\hbox{*}}{\eightrm Supported in part
       by the Robert A. Welch Foundation and N.S.F. Grants 
       PHY-880637 and\break PHY-8605978.}}}

\def\newpage{\vfill\eject}
\def\abstract#1{\centerline{\bf ABSTRACT}\vskip.2truein{\narrower\noindent#1
       \smallskip}}

\def\runninghead#1#2{\voffset=2\baselineskip\nopagenumbers
       \headline={\ifodd\pageno\rightheadline\else \leftheadline\fi}
       \def\rightheadline{{\sl#1}\hfill{\rm\folio}}
       \def\leftheadline{{\rm\folio}\hfill{\sl#2}}}
\def\SS{\mathhexbox278}

\newcount\footnoteno
\def\Footnote#1{\advance\footnoteno by 1
                \let\SF=\empty 
                \ifhmode\edef\SF{\spacefactor=\the\spacefactor}\/\fi
                $^{\the\footnoteno}$\ignorespaces
                \SF\vfootnote{$^{\the\footnoteno}$}{#1}}

\def\figbox#1#2#3{\vbox{\vskip15pt
                   \vbox{\hrule
                    \hbox{\vrule
                     \vbox{\vskip12truept\centerline #1 \vskip6truept
                          {\hskip.4truein\vbox{\hsize=5truein\noindent
                          {\bf Figure\hskip5truept#2:}\hskip5truept#3}}
                     \vskip18truept}
                    \vrule}
                   \hrule}}}
\def\place#1#2#3{\vbox to0pt{\kern-\parskip\kern-7pt
                             \kern-#2truein\hbox{\kern#1truein #3}
                             \vss}\nointerlineskip}
\def\figurecaption#1#2{\kern.75truein\vbox{\hsize=5truein\noindent{\bf Figure
    \figlabel{#1}:} #2}}
\def\tablecaption#1#2{\kern.75truein\lower12truept\hbox{\vbox{\hsize=5truein
    \noindent{\bf Table\hskip5truept\tablabel{#1}:} #2}}}
\def\boxed#1{\lower3pt\hbox{
                       \vbox{\hrule\hbox{\vrule
                         \vbox{\kern2pt\hbox{\kern3pt#1\kern3pt}\kern3pt}\vrule}
                         \hrule}}}
%%%%%%%%%%%%%%%%%%%%%%%%%%%%%%%%%%%%%%%%%%%%%%%%%%%%%%%%%%%%%%%%%%%%%%%%%%%%%%%
%Greek characters%%%%%%%%%%%%%%%%%%%%%%%%%%%%%%%%%%%%%%%%%%%%%%%%%%%%%%%%%%%%%%
%%%%%%%%%%%%%%%%%%%%%%%%%%%%%%%%%%%%%%%%%%%%%%%%%%%%%%%%%%%%%%%%%%%%%%%%%%%%%%%
\def\a{\alpha}

\def\g{\gamma}\def\G{\Gamma}
\def\d{\delta}\def\D{\Delta}
\def\e{\epsilon}

\def\th{\theta}

\def\l{\lambda}\def\L{\Lambda}
\def\m{\mu}

\def\x{\xi}

\def\p{\pi}

\def\S{\Sigma}
\def\t{\tau}

\def\ch{\chi}

%%%%%%%%%%%%%%%%%%%%%%%%%%%%%%%%%%%%%%%%%%%%%%%%%%%%%%%%%%%%%%%%%%%%%%%%%%%%%%%
%Calligraphic capitals%%%%%%%%%%%%%%%%%%%%%%%%%%%%%%%%%%%%%%%%%%%%%%%%%%%%%%%%%
%%%%%%%%%%%%%%%%%%%%%%%%%%%%%%%%%%%%%%%%%%%%%%%%%%%%%%%%%%%%%%%%%%%%%%%%%%%%%%%
\def\ca#1{\relax\ifmmode {{\cal #1}}\else $\cal #1$\fi}

\def\calb{{\cal B}}

\def\calm{{\cal M}}

%%%%%%%%%%%%%%%%%%%%%%%%%%%%%%%%%%%%%%%%%%%%%%%%%%%%%%%%%%%%%%%%%%%%%%%%%%%%%%%
% Poor man's Blackboard Bold%%%%%%%%%%%%%%%%%%%%%%%%%%%%%%%%%%%%%%%%%%%%%%%%%%%
%%%%%%%%%%%%%%%%%%%%%%%%%%%%%%%%%%%%%%%%%%%%%%%%%%%%%%%%%%%%%%%%%%%%%%%%%%%%%%%
\def\inbar{\vrule height1.5ex width.4pt depth0pt}
\def\IB{\relax{\rm I\kern-.18em B}}
\def\IC{\relax\hbox{\kern.25em$\inbar\kern-.3em{\rm C}$}}
\def\ID{\relax{\rm I\kern-.18em D}}
\def\IE{\relax{\rm I\kern-.18em E}}
\def\IF{\relax{\rm I\kern-.18em F}}
\def\IG{\relax\hbox{\kern.25em$\inbar\kern-.3em{\rm G}$}}
\def\IH{\relax{\rm I\kern-.18em H}}
\def\II{\relax{\rm I\kern-.18em I}}
\def\IK{\relax{\rm I\kern-.18em K}}
\def\IL{\relax{\rm I\kern-.18em L}}
\def\IM{\relax{\rm I\kern-.18em M}}
\def\IN{\relax{\rm I\kern-.18em N}}
\def\IO{\relax\hbox{\kern.25em$\inbar\kern-.3em{\rm O}$}}
\def\IP{\relax{\rm I\kern-.18em P}}
\def\IQ{\relax\hbox{\kern.25em$\inbar\kern-.3em{\rm Q}$}}
\def\IR{\relax{\rm I\kern-.18em R}}
\def\IZ{\relax\ifmmode\hbox{\ss Z\kern-.4em Z}\else{\ss Z\kern-.4em Z}\fi}
\def\IGa{\relax{\rm I}\kern-.18em\Gamma}
\def\IPi{\relax{\rm I}\kern-.18em\Pi}
\def\ITh{\relax\hbox{\kern.25em$\inbar\kern-.3em\Theta$}}
\def\IOm{\relax\thinspace\inbar\kern1.95pt\inbar\kern-5.525pt\Omega}

%Papers, Lecture Notes on Complex Manifolds etc.

\def\ie{{\it i.e.,\ \/}}

\def\noblackboxes{\overfullrule=0pt}

\def\cy{Calabi--Yau} 
\def\cym{Calabi--Yau manifold}
\def\cys{Calabi--Yau manifolds}

\def\K{K\"ahler}

\def\H#1#2{\relax\ifmmode {H^{#1#2}}\else $H^{#1 #2}$\fi}
\def\M{\relax\ifmmode{\calm}\else $\calm$\fi}

\def\Bigcheck{\lower3.8pt\hbox{\smash{\hbox{{\twentyonerm \v{}}}}}}
\def\bigboldcheck{\smash{\hbox{{\seventeenbold\v{}}}}}

\def\Bighat{\lower3.8pt\hbox{\smash{\hbox{{\twentyonerm \^{}}}}}}

\def\Msharp{\relax\ifmmode{\calm^\sharp}\else $\smash{\calm^\sharp}$\fi}
\def\Mflat{\relax\ifmmode{\calm^\flat}\else $\smash{\calm^\flat}$\fi}
\def\preMcheck{\kern2pt\hbox{\Bigcheck\kern-12pt{$\cal M$}}}
\def\Mcheck{\relax\ifmmode\preMcheck\else $\preMcheck$\fi}
\def\preMhat{\kern2pt\hbox{\Bighat\kern-12pt{$\cal M$}}}
\def\Mhat{\relax\ifmmode\preMhat\else $\preMhat$\fi}

\def\Bsharp{\relax\ifmmode{\calb^\sharp}\else $\calb^\sharp$\fi}
\def\Bflat{\relax\ifmmode{\calb^\flat}\else $\calb^\flat$ \fi}
\def\preBcheck{\hbox{\Bigcheck\kern-9pt{$\cal B$}}}
\def\Bcheck{\relax\ifmmode\preBcheck\else $\preBcheck$\fi}
\def\preBhat{\hbox{\Bighat\kern-9pt{$\cal B$}}}
\def\Bhat{\relax\ifmmode\preBhat\else $\preBhat$\fi}

\def\figBcheck{\kern3pt\hbox{\raise1pt\hbox{\bigboldcheck}\kern-11pt
    {\twelvecal B}}}
\def\figBsharp{{\twelvecal B}\raise5pt\hbox{$\twelvemath\sharp$}}
\def\figBflat{{\twelvecal B}\raise5pt\hbox{$\twelvemath\flat$}}

\def\gcheck{\hbox{\lower2.5pt\hbox{\Bigcheck}\kern-8pt$\g$}}
\def\lhat{\hbox{\raise.5pt\hbox{\Bighat}\kern-8pt$\l$}}

\def\Fcheck{\kern2pt\hbox{\raise1pt\hbox{\Bigcheck}\kern-10pt{$\cal F$}}}
\def\Fhat{\kern2pt\hbox{\raise1pt\hbox{\Bighat}\kern-10pt{$\cal F$}}}
 
\def\cp#1{\relax\ifmmode {\IP\kern-2pt{}_{#1}}\else $\IP\kern-2pt{}_{#1}$\fi}
\def\h#1#2{\relax\ifmmode {b_{#1#2}}\else $b_{#1#2}$\fi}

\def\half{{1\over 2}}

\def\frac#1#2{{#1\over #2}}
\def\vol{g^\half d^6x}

\def\cone{\relax\thinspace\hbox{$<\kern-.8em{)}$}}
\mathchardef\mho"0A30

\def\asymp{\sim}
\def\-{\hphantom{-}}

%References

\def\npb#1{Nucl.\ Phys.\ {\bf B#1}}

% Pictures

\def\picture #1 by #2 (#3){\vbox to #2{\hrule width #1 height 0pt depth 0pt
                                       \vfill\special{picture #3}}}
\def\scaledpicture #1 by #2 (#3 scaled #4){{\dimen0=#1 \dimen1=#2
           \divide\dimen0 by 1000 \multiply\dimen0 by #4
            \divide\dimen1 by 1000 \multiply\dimen1 by #4
            \picture \dimen0 by \dimen1 (#3 scaled #4)}}
\def\illustration #1 by #2 (#3){\vbox to #2{\hrule width #1 height 0pt depth 0pt
                                       \vfill\special{illustration #3}}}
\def\scaledillustration #1 by #2 (#3 scaled #4){{\dimen0=#1 \dimen1=#2
           \divide\dimen0 by 1000 \multiply\dimen0 by #4
            \divide\dimen1 by 1000 \multiply\dimen1 by #4
            \illustration \dimen0 by \dimen1 (#3 scaled #4)}}

%Letters, Letters of Recommendation, Referee's Reports, Itineraries, etc.

\def\delaOssa{\nobreak\vskip1truein\hbox to\hsize
       {\hskip 4truein Xenia de la Ossa\hfill}}

\def\hoy{\number\day\space de \ifcase\month\or enero\or febrero\or marzo\or
       abril\or mayo\or junio\or julio\or agosto\or septiembre\or octubre\or
       noviembre\or diciembre\fi\space de \number\year}

%%%%%%%%%%%%%%%%%%%%%%%%%%%%%%%%%%%%%%%%%%%%%%%%%%%%%%%%%%%%%%%%%%%%%%%%%%%%%%%
%Equation macros%%%%%%%%%%%%%%%%%%%%%%%%%%%%%%%%%%%%%%%%%%%%%%%%%%%%%%%%%%%%%%%
%%%%%%%%%%%%%%%%%%%%%%%%%%%%%%%%%%%%%%%%%%%%%%%%%%%%%%%%%%%%%%%%%%%%%%%%%%%%%%%
\def\cropen#1{\crcr\noalign{\vskip #1}}%For use with \eqalign

\newif\ifproofmode
\proofmodefalse

\newif\ifforwardreference
\forwardreferencefalse

\newif\ifchapternumbers
\chapternumbersfalse

\newif\ifcontinuousnumbering
\continuousnumberingfalse

\newif\iffigurechapternumbers
\figurechapternumbersfalse

\newif\ifcontinuousfigurenumbering
\continuousfigurenumberingfalse

\newif\iftablechapternumbers
\tablechapternumbersfalse

\newif\ifcontinuoustablenumbering
\continuoustablenumberingfalse

\font\eqsixrm=cmr6

\def\marginstyle{\eqsixrm}

\newtoks\chapletter
\newcount\chapno
\newcount\eqlabelno
\newcount\figureno
\newcount\tableno

\chapno=0
\eqlabelno=0
\figureno=0
\tableno=0

\def\chapfolio{\ifnum\chapno>0 \the\chapno\else\the\chapletter\fi}

\def\bumpchapno{\ifnum\chapno>-1 \global\advance\chapno by 1
\else\global\advance\chapno by -1 \setletter\chapno\fi
\ifcontinuousnumbering\else\global\eqlabelno=0 \fi
\ifcontinuousfigurenumbering\else\global\figureno=0 \fi
\ifcontinuoustablenumbering\else\global\tableno=0 \fi}

\def\setletter#1{\ifcase-#1{}\or{}%
\or\global\chapletter={A}%
\or\global\chapletter={B}%
\or\global\chapletter={C}%
\or\global\chapletter={D}%
\or\global\chapletter={E}%
\or\global\chapletter={F}%
\or\global\chapletter={G}%
\or\global\chapletter={H}%
\or\global\chapletter={I}%
\or\global\chapletter={J}%
\or\global\chapletter={K}%
\or\global\chapletter={L}%
\or\global\chapletter={M}%
\or\global\chapletter={N}%
\or\global\chapletter={O}%
\or\global\chapletter={P}%
\or\global\chapletter={Q}%
\or\global\chapletter={R}%
\or\global\chapletter={S}%
\or\global\chapletter={T}%
\or\global\chapletter={U}%
\or\global\chapletter={V}%
\or\global\chapletter={W}%
\or\global\chapletter={X}%
\or\global\chapletter={Y}%
\or\global\chapletter={Z}\fi}

\def\tempsetletter#1{\ifcase-#1{}\or{}%
\or\global\chapletter={A}%
\or\global\chapletter={B}%
\or\global\chapletter={C}%
\or\global\chapletter={D}%
\or\global\chapletter={E}%
\or\global\chapletter={F}%
\or\global\chapletter={G}%
\or\global\chapletter={H}%
\or\global\chapletter={I}%
\or\global\chapletter={J}%
\or\global\chapletter={K}%
\or\global\chapletter={L}%
\or\global\chapletter={M}%
\or\global\chapletter={N}%
\or\global\chapletter={O}%
\or\global\chapletter={P}%
\or\global\chapletter={Q}%
\or\global\chapletter={R}%
\or\global\chapletter={S}%
\or\global\chapletter={T}%
\or\global\chapletter={U}%
\or\global\chapletter={V}%
\or\global\chapletter={W}%
\or\global\chapletter={X}%
\or\global\chapletter={Y}%
\or\global\chapletter={Z}\fi}

\def\chapshow#1{\ifnum#1>0 \relax#1%
\else{\tempsetletter{\number#1}\chapno=#1\chapfolio}\fi}

\def\chaplabel#1{\bumpchapno\ifproofmode\ifforwardreference
\immediate\write1{\noexpand\expandafter\noexpand\def
\noexpand\csname CHAPLABEL#1\endcsname{\the\chapno}}\fi\fi
\global\expandafter\edef\csname CHAPLABEL#1\endcsname
{\the\chapno}\ifproofmode\llap{\hbox{\marginstyle #1\ }}\fi\chapfolio}

\def\chapref#1{\ifundefined{CHAPLABEL#1}??\ifproofmode\ifforwardreference%
\else\write16{ ***Undefined Chapter Reference #1*** }\fi
\else\write16{ ***Undefined Chapter Reference #1*** }\fi
\else\edef\LABxx{\getlabel{CHAPLABEL#1}}\chapshow\LABxx\fi
\ifproofmode\write2{Chapter #1}\fi}

\def\eqnum{\global\advance\eqlabelno by 1
\eqno(\ifchapternumbers\chapfolio.\fi\the\eqlabelno)}

\def\eqlabel#1{\global\advance\eqlabelno by 1 \ifproofmode\ifforwardreference
\immediate\write1{\noexpand\expandafter\noexpand\def
\noexpand\csname EQLABEL#1\endcsname{\the\chapno.\the\eqlabelno?}}\fi\fi
\global\expandafter\edef\csname EQLABEL#1\endcsname
{\the\chapno.\the\eqlabelno?}\eqno(\ifchapternumbers\chapfolio.\fi
\the\eqlabelno)\ifproofmode\rlap{\hbox{\marginstyle #1}}\fi}

\def\eqalignnum{\global\advance\eqlabelno by 1
&(\ifchapternumbers\chapfolio.\fi\the\eqlabelno)}

\def\eqalignlabel#1{\global\advance\eqlabelno by 1 \ifproofmode 
\ifforwardreference\immediate\write1{\noexpand\expandafter\noexpand\def
\noexpand\csname EQLABEL#1\endcsname{\the\chapno.\the\eqlabelno?}}\fi\fi
\global\expandafter\edef\csname EQLABEL#1\endcsname
{\the\chapno.\the\eqlabelno?}&(\ifchapternumbers\chapfolio.\fi
\the\eqlabelno)\ifproofmode\rlap{\hbox{\marginstyle #1}}\fi}

\def\eqref#1{\hbox{(\ifundefined{EQLABEL#1}***)\ifproofmode\ifforwardreference%
\else\write16{ ***Undefined Equation Reference #1*** }\fi
\else\write16{ ***Undefined Equation Reference #1*** }\fi
\else\edef\LABxx{\getlabel{EQLABEL#1}}%
\def\LAByy{\expandafter\stripchap\LABxx}\ifchapternumbers%
\chapshow{\LAByy}.\expandafter\stripeq\LABxx%
\else\ifnum\number\LAByy=\chapno\relax\expandafter\stripeq\LABxx%
\else\chapshow{\LAByy}.\expandafter\stripeq\LABxx\fi\fi)\fi}%
\ifproofmode\write2{Equation #1}\fi}

\def\fignum{\global\advance\figureno by 1
\relax\iffigurechapternumbers\chapfolio.\fi\the\figureno}

\def\figlabel#1{\global\advance\figureno by 1
\relax\ifproofmode\ifforwardreference
\immediate\write1{\noexpand\expandafter\noexpand\def
\noexpand\csname FIGLABEL#1\endcsname{\the\chapno.\the\figureno?}}\fi\fi
\global\expandafter\edef\csname FIGLABEL#1\endcsname
{\the\chapno.\the\figureno?}\iffigurechapternumbers\chapfolio.\fi
\ifproofmode\llap{\hbox{\marginstyle#1
\kern1.2truein}}\relax\fi\the\figureno}

\def\figref#1{\hbox{\ifundefined{FIGLABEL#1}!!!!\ifproofmode\ifforwardreference%
\else\write16{ ***Undefined Figure Reference #1*** }\fi
\else\write16{ ***Undefined Figure Reference #1*** }\fi
\else\edef\LABxx{\getlabel{FIGLABEL#1}}%
\def\LAByy{\expandafter\stripchap\LABxx}\iffigurechapternumbers%
\chapshow{\LAByy}.\expandafter\stripeq\LABxx%
\else\ifnum \number\LAByy=\chapno\relax\expandafter\stripeq\LABxx%
\else\chapshow{\LAByy}.\expandafter\stripeq\LABxx\fi\fi\fi}%
\ifproofmode\write2{Figure #1}\fi}

\def\tabnum{\global\advance\tableno by 1
\relax\iftablechapternumbers\chapfolio.\fi\the\tableno}

\def\tablabel#1{\global\advance\tableno by 1
\relax\ifproofmode\ifforwardreference
\immediate\write1{\noexpand\expandafter\noexpand\def
\noexpand\csname TABLABEL#1\endcsname{\the\chapno.\the\tableno?}}\fi\fi
\global\expandafter\edef\csname TABLABEL#1\endcsname
{\the\chapno.\the\tableno?}\iftablechapternumbers\chapfolio.\fi
\ifproofmode\llap{\hbox{\marginstyle#1
\kern1.2truein}}\relax\fi\the\tableno}

\def\tabref#1{\hbox{\ifundefined{TABLABEL#1}!!!!\ifproofmode\ifforwardreference%
\else\write16{ ***Undefined Table Reference #1*** }\fi
\else\write16{ ***Undefined Table Reference #1*** }\fi
\else\edef\LABtt{\getlabel{TABLABEL#1}}%
\def\LABTT{\expandafter\stripchap\LABtt}\iftablechapternumbers%
\chapshow{\LABTT}.\expandafter\stripeq\LABtt%
\else\ifnum\number\LABTT=\chapno\relax\expandafter\stripeq\LABtt%
\else\chapshow{\LABTT}.\expandafter\stripeq\LABtt\fi\fi\fi}%
\ifproofmode\write2{Table#1}\fi}

\def\fig{Figure~}

\newdimen\sectionskip     \sectionskip=20truept
\newcount\sectno
\def\section#1#2{\sectno=0 \null\vskip\sectionskip
    \centerline{\chaplabel{#1}.~~{\bf#2}}\nobreak\vskip.2truein
    \noindent\ignorespaces}

\def\advancesectno{\global\advance\sectno by 1}
\def\sectfolio{\number\sectno}
\def\subsection#1{\goodbreak\advancesectno\null\vskip10pt
                  \noindent\chapfolio.~\sectfolio.~{\bf #1}
                  \nobreak\vskip.05truein\noindent\ignorespaces}

\def\uttg#1{\null\vskip.1truein
    \ifproofmode \line{\hfill{\bf Draft}:
    UTTG--{#1}--\number\year}\line{\hfill\today}
    \else \line{\hfill UTTG--{#1}--\number\year}
    \line{\hfill\ifcase\month\or January\or February\or March\or April\or May\or June
    \or July\or August\or September\or October\or November\or December\fi
    \space\number\year}\fi}

\def\contents{\noindent
   {\bf Contents\Z}\nobreak\vskip.05truein\noindent\ignorespaces}

\def\getlabel#1{\csname#1\endcsname}
\def\ifundefined#1{\expandafter\ifx\csname#1\endcsname\relax}
\def\stripchap#1.#2?{#1}
\def\stripeq#1.#2?{#2}

%%%%%%%%%%%%%%%%%%%%%%%%%%%%%%%%%%%%%%%%%%%%%%%%%%%%%%%%%%%%%%%%%%%%%%%%%%%%%%
%Reference macros%%%%%%%%%%%%%%%%%%%%%%%%%%%%%%%%%%%%%%%%%%%%%%%%%%%%%%%%%%%%%
%%%%%%%%%%%%%%%%%%%%%%%%%%%%%%%%%%%%%%%%%%%%%%%%%%%%%%%%%%%%%%%%%%%%%%%%%%%%%%
%
\catcode`@=11 % This allows us to modify PLAIN macros.
\def\space@ver#1{\let\@sf=\empty\ifmmode#1\else\ifhmode%
\edef\@sf{\spacefactor=\the\spacefactor}\unskip${}#1$\relax\fi\fi}
\newcount\referencecount     \referencecount=0
\newif\ifreferenceopen       \newwrite\referencewrite
\newtoks\rw@toks
\def\refmark#1{\relax[#1]}
\def\refend{\refmark{\number\referencecount}}
\newcount\lastrefsbegincount \lastrefsbegincount=0
\def\refsend{\refmark{\count255=\referencecount%
\advance\count255 by -\lastrefsbegincount%
\ifcase\count255 \number\referencecount%
\or\number\lastrefsbegincount,\number\referencecount%
\else\number\lastrefsbegincount-\number\referencecount\fi}}
\def\refch@ck{\chardef\rw@write=\referencewrite
\ifreferenceopen\else\referenceopentrue
\immediate\openout\referencewrite=referenc.texauxil \fi}
%
% In \obeyendofline, we say `\let^^M=\relax
{\catcode`\^^M=\active % these lines must end with %
  \gdef\obeyendofline{\catcode`\^^M\active \let^^M\ }}%
%
% In \ignoreendofline, we say `\let^^M=\relax
{\catcode`\^^M=\active % these lines must end with %
  \gdef\ignoreendofline{\catcode`\^^M=5}}
{\obeyendofline\gdef\rw@start#1{\def\t@st{#1}\ifx\t@st\blankend%
\endgroup\@sf\relax\else\ifx\t@st\bl@nkend\endgroup\@sf\relax%
\else\rw@begin#1
\backtotext
\fi\fi}}
{\obeyendofline\gdef\rw@begin#1
{\def\n@xt{#1}\rw@toks={#1}\relax%
\rw@next}}
\def\blankend{}
{\obeylines\gdef\bl@nkend{
}}
\newif\iffirstrefline  \firstreflinetrue
\def\rwr@teswitch{\ifx\n@xt\blankend\let\n@xt=\rw@begin%
\else\iffirstrefline\global\firstreflinefalse%
\immediate\write\rw@write{\noexpand\obeyendofline\the\rw@toks}%
\let\n@xt=\rw@begin%
\else\ifx\n@xt\rw@@d \def\n@xt{\immediate\write\rw@write{%
\noexpand\ignoreendofline}\endgroup\@sf}%
\else\immediate\write\rw@write{\the\rw@toks}%
\let\n@xt=\rw@begin\fi\fi\fi}
\def\rw@next{\rwr@teswitch\n@xt}
\def\rw@@d{\backtotext} \let\rw@end=\relax
\let\backtotext=\relax

\newdimen\refindent     \refindent=30pt
\def\Textindent#1{\noindent\llap{#1\enspace}\ignorespaces}
\def\refitem#1{\par\hangafter=0 \hangindent=\refindent\Textindent{#1}}
\def\REFNUM#1{\space@ver{}\refch@ck\firstreflinetrue%
\global\advance\referencecount by 1 \xdef#1{\the\referencecount}}
\def\refnum#1{\space@ver{}\refch@ck\firstreflinetrue%
\global\advance\referencecount by 1\xdef#1{\the\referencecount}\refend}

\def\REF#1{\REFNUM#1%
\immediate\write\referencewrite{%
\noexpand\refitem{#1.}}%
\begingroup\obeyendofline\rw@start}
\def\ref{\refnum\?%
\immediate\write\referencewrite{\noexpand\refitem{\?.}}%
\begingroup\obeyendofline\rw@start}
\def\Ref#1{\refnum#1%
\immediate\write\referencewrite{\noexpand\refitem{#1.}}%
\begingroup\obeyendofline\rw@start}
\def\REFS#1{\REFNUM#1\global\lastrefsbegincount=\referencecount%
\immediate\write\referencewrite{\noexpand\refitem{#1.}}%
\begingroup\obeyendofline\rw@start}

\def\REFSCON#1{\REF#1}

\def\cite#1{\refmark#1}
\catcode`@=12 % at signs are no longer letters
%
%%%%%%%%%%%%%%%%%%%%%%%%%%%%%%%%%%%%%%%%%%%%%%%%%%%%%%%%%%%%%%%%%%%%%%%%%%%%%%
%%%%%%%%%%%%%%%%%%%%%%%%%%%%[ Epsf macro    ]%%%%%%%%%%%%%%%%%%%%%%%%%%%%%%%%%
%%%%%%%%%%%%%%%%%%%%%%%%%%%%%%%%%%%%%%%%%%%%%%%%%%%%%%%%%%%%%%%%%%%%%%%%%%%%%%
%
\input epsf.tex
%
%%%%%%%%%%%%%%%%%%%%%%%%%%%%%%%%%%%%%%%%%%%%%%%%%%%%%%%%%%%%%%%%%%%%%%%%%%%%%%
%%%%%%%%%%%%%%%%%%%%%%%%%%[  Set macro flags  ]%%%%%%%%%%%%%%%%%%%%%%%%%%%%%%%
%%%%%%%%%%%%%%%%%%%%%%%%%%%%%%%%%%%%%%%%%%%%%%%%%%%%%%%%%%%%%%%%%%%%%%%%%%%%%%
%
\proofmodefalse
\baselineskip=15pt plus 1pt minus 1pt
\parskip=5pt
\chapternumberstrue
\forwardreferencefalse
\figurechapternumberstrue
\tablechapternumberstrue
\ifproofmode
\immediate\openout2=allcrossreferfile \fi
\ifforwardreference\input labelfile
\ifproofmode\immediate\openout1=labelfile \fi\fi
\noblackboxes
\hfuzz=1pt
\vfuzz=2pt
%
%%%%%%%%%%%%%%%%%%%%%%%%%%%%%%%%%%%%%%%%%%%%%%%%%%%%%%%%%%%%%%%%%%%%%%%%%%%%%%
%%%%%%%%%%%%%%%%%[ Some more or less useful definitions ]%%%%%%%%%%%%%%%%%%%%%
%%%%%%%%%%%%%%%%%%%%%%%%%%%%%%%%%%%%%%%%%%%%%%%%%%%%%%%%%%%%%%%%%%%%%%%%%%%%%%
%
\def\hourandminute{\count255=\time\divide\count255 by 60
\xdef\hour{\number\count255}
\multiply\count255 by -60\advance\count255 by\time
\hour:\ifnum\count255<10 0\fi\the\count255}

\def\chaplabel#1{\bumpchapno\ifproofmode\ifforwardreference
\immediate\write1{\noexpand\expandafter\noexpand\def
\noexpand\csname CHAPLABEL#1\endcsname{\the\chapno}}\fi\fi
\global\expandafter\edef\csname CHAPLABEL#1\endcsname
{\the\chapno}\ifproofmode
\llap{\hbox{\marginstyle #1\ifnum\chapno > -1\ \else
\hskip1.3truein\fi}}\fi\chapfolio}

\def\section#1#2{\sectno=0 \null\vskip\sectionskip
    \ifnum\chapno > -1 
    \centerline{\fourteenrm\chaplabel{#1}.~~\fourteenbold#2}
    \else
    \centerline{\fourteenbold Appendix\ \chaplabel{#1}: {#2}}\fi
    \nobreak\vskip.2truein
    \noindent\ignorespaces}
\def\subsection#1{\goodbreak\advancesectno\null\vskip10pt
                  \noindent{\it \chapfolio.\sectfolio.~#1}
                  \nobreak\vskip.05truein\noindent\ignorespaces}
\def\subsubsection#1{\goodbreak
                  \noindent$\underline{\hbox{#1}}$
                  \nobreak\vskip-5pt\noindent\ignorespaces}
\def\cite#1{\refmark{#1}}
\def\\{\hfill\break}
\def\cropen#1{\crcr\noalign{\vskip #1}}%For use with \eqalign
\def\contents{\line{{\fourteenbold Contents}\hfill}\nobreak\vskip.05truein\noindent%
              \ignorespaces}

\def\titlebox#1#2{\lower7pt\hbox{%
\hsize=.75in\vbox{\vskip5pt\centerline{#1}\vskip5pt\centerline{#2}}}}

\def\F{\hbox{F}}
\def\Het{\hbox{Het}}
\def\IIB{\hbox{IIB}}

\def\pts{\hbox{pts}}
\def\inte{\hbox{int}}
\def\vol{\hbox{vol}}
\def\codim{\hbox{\eightrm codim}}
\def\dim{\hbox{\eightrm dim}}

\def\lra{\longrightarrow}
\def\lla{\longleftarrow}

%
%%%%%%%%%%%%%%%%%%%%%%%%%%%%%%%%%%%%%%%%%%%%%%%%%%%%%%%%%%%%%%%%%%%%%%%%%%%%%%
%%%%%%%%%%%%%%%%%%%%%%%%%%%%[ The Title    ]%%%%%%%%%%%%%%%%%%%%%%%%%%%%%%%%%%
%%%%%%%%%%%%%%%%%%%%%%%%%%%%%%%%%%%%%%%%%%%%%%%%%%%%%%%%%%%%%%%%%%%%%%%%%%%%%%
%
\nopagenumbers\pageno=0
\null\vskip-30pt
\vfootnote{\eightrm *}{\eightrm Supported in part
       by the Robert A. Welch Foundation, N.S.F. grants
       DMS9706707 and PHY-9511632.}
\vskip.3in
\vbox{\baselineskip=12pt
\rightline{\eightrm UTTG-06-99}\vskip-3pt
\rightline{\eightrm hep-th/0001208}\vskip-3pt
\rightline{\eightrm 29 January 2000}
\vskip.25truein
\centerline{\seventeenrm Toric Calabi-Yau Fourfolds}
\vskip12pt
\centerline{\seventeenrm Duality Between N=1 Theories and}
\vskip2pt
\centerline{\seventeenrm \hphantom{\ \raise6pt\hbox{*}}% 
Divisors that Contribute to the Superpotential\ \raise6pt\hbox{*}}
\vskip.4truein
\centerline{%
      {\csc \hphantom{$^{1,3}$}Volker~Braun}$^{1,3}$,\quad
      {\csc Philip~Candelas}$^{2,3}$,\quad
      {\csc Xenia~de~la~Ossa}$^{2,3}$}
\centerline{\csc and}
\centerline{\csc \hphantom{$^4$}Antonella Grassi$^4$}
\vskip.1truein\bigskip
\centerline{
\vtop{\hsize = 3.0truein
\centerline{$^1$\it Institut f\"ur Physik}
\centerline{\it Humboldt Universit\"at}
\centerline{\it Invalidenstrasse 110,}
\centerline{\it 10115 Berlin, Germany}}\hfil
\vtop{\hsize = 3.0truein
\centerline{$^2$\it Mathematical Institute}
\centerline{\it Oxford University}
\centerline{\it 24-29 St.\ Giles'}
\centerline{\it Oxford OX1 3LB, England}}}
\vskip.2in
\centerline{
\vtop{\hsize = 3.0truein
\centerline{$^3$\it Theory Group}
\centerline{\it Department of Physics}
\centerline{\it University of Texas}
\centerline{\it Austin, TX 78712, USA}}\
\vtop{\hsize = 3.0truein
\centerline{$^4$\it Department of Mathematics}
\centerline{\it 209 South 33rd St.}
\centerline{\it University of Pennsylvania}
\centerline{\it Philadelphia, PA 19104, USA}}}
\vskip.2in\bigskip
\centerline{\bf ABSTRACT}
\vskip.1truein 
\noindent We study issues related to $F$-theory on
Calabi-Yau fourfolds and its  duality to heterotic theory for \cy\
threefolds. We discuss principally fourfolds that are described by
reflexive polyhedra and show how to read off some of the data for the
heterotic theory from the polyhedron. We give a procedure for
constructing examples with given gauge groups and describe some
of these examples in detail. Interesting features arise when the local pieces are
fitted into a global manifold. An important issue is how to compute
the superpotential explicitly.  Witten has shown that the condition for a
divisor to contribute to the superpotential is that it have arithmetic
genus 1. Divisors associated with the short roots of non-simply laced
gauge groups do not always satisfy this condition while the divisors
associated to all other roots do. For such a `dissident' divisor we 
distinguish cases for which 
$\chi ({\cal O}_D) > 1$ corresponding to an $X$ that is not general in
moduli (in the toric case this corresponds to the existence of non-toric
parameters). In these cases the `dissident'  divisor $D$ does not remain
an effective divisor for general complex structure.  If however
$\chi ( {\cal O}_D) \leq 0$, then the divisor is general in moduli 
and there is a genuine instability.}
%
%%%%%%%%%%%%%%%%%%%%%%%%%%%%%%%%%%%%%%%%%%%%%%%%%%%%%%%%%%%%%%%%%%%%%%%%%%%%%%
%%%%%%%%%%%%%%%%%%%%%%%%%%%%[   Contents   ]%%%%%%%%%%%%%%%%%%%%%%%%%%%%%%%%%%
%%%%%%%%%%%%%%%%%%%%%%%%%%%%%%%%%%%%%%%%%%%%%%%%%%%%%%%%%%%%%%%%%%%%%%%%%%%%%%
% 
\newpage
\vbox{\baselineskip=12pt
\contents
\vskip10pt
\item{1.~}Introduction
\vskip5pt
\item{2.~}Toric Preliminaries
\vskip5pt
\item{3.~}Heterotic Structure
\itemitem{\it 3.1~}{\it The Fibrations}
\vskip5pt
\item{4.~}Construction of $X$
\itemitem{\it 4.1~}{\it The KLRY Spaces}
\itemitem{\it 4.2~}{\it Constructing the Fan for $\widetilde{X}$}
\itemitem{\it 4.3~}{\it Extending the Groups}
\itemitem{\it 4.4~}{\it New groups}
\vskip5pt
\item{5.~}The Yukawa Couplings and Mori Cone
\itemitem{\it 5.1~}{\it the Yukawa Couplings}
\itemitem{\it 5.2~}{\it The Mori Cone of $X$}
\itemitem{\it 5.3~}{\it Volumes of the Divisors}
\vskip5pt
\item{6.~}The Superpotential
\itemitem{\it 6.1~}{\it Characterization of the Divisors
Contributing to the Worldsheet Instantons}
\itemitem{\it 6.2~}{\it Comparison with the Heterotic Superpotential}
\itemitem{\it 6.3~}{\it Andreas' Examples}
\itemitem{\it 6.4~}{\it Comparison with $d=3$ Dimensional Yang-Mills Theory}
\itemitem{\it 6.5~}{\it New Features for a Non-Simply Laced Group}
\vskip5pt
\item{A.~}Appendix: Geometry of $Z$
\itemitem{\it A.1~}{\it The Divisors}
\itemitem{\it A.2~}{\it Projection to $B^Z$}
\vskip5pt
\item{B.~}Appendix: The Divisors for the Spaces $Y^\pm$
}
\newpage
 %
%%%%%%%%%%%%%%%%%%%%%%%%%%%%%%%%%%%%%%%%%%%%%%%%%%%%%%%%%%%%%%%%%%%%%%%%%%%%%%
%%%%%%%%%%%%%%%%%%%%%%%%%%%%[ The Article! ]%%%%%%%%%%%%%%%%%%%%%%%%%%%%%%%%%%
%%%%%%%%%%%%%%%%%%%%%%%%%%%%%%%%%%%%%%%%%%%%%%%%%%%%%%%%%%%%%%%%%%%%%%%%%%%%%%
 %
\headline={\ifproofmode\hfil\eightrm draft:\ \today\
\hourandminute\else\hfil\fi}
\footline={\rm\hfil\folio\hfil}
\pageno=1
\section{intro}{Introduction}
This paper is devoted to a number of issues pertaining to the
compactification of $\F$-theory on \cy\ fourfolds.  For fourfolds, $X$, of particular
structure there are believed to be interesting dualities~
\Ref{\VafaF}{C. Vafa, ``Evidence for $F$-Theory'', 
\npb{469} (1996) 403, hep-th/9602022.}
$$
\F[X] = \IIB[B^X] = \Het[Z^X, V^X]~, \eqlabel{duals}$$
that relate $\F$-theory on $X$ to $\IIB$ string theory compactified on a
(non \cy) threefold $B^X$ and also to a heterotic compactification on a \cy\
threefold $Z^X$ with vector bundle $V^X$. The relation between heterotic 
compactification on threefolds and \cy\ fourfolds that these dualities entail is
particularly interesting since it offers the hope of insight into the important
but so far poorly understood $(0,2)$ compactifications of the heterotic string. Relation
\eqref{duals} suggests that heterotic theories on \cy\ threefolds are, in some sense,
classified by \cy\ fourfolds. We will here concern ourselves largely with \cy\ fourfolds
that are themselves described by reflexive polyhedra. Thus a class of heterotic
theories on \cy\ threefolds are described by reflexive five dimensional polyhedra. The
question that we seek to answer is to what extent we can read off the data for the
heterotic theory from this polyhedron. 

A basic question is how to read off the the threefold of the heterotic theory from the data.
This is accomplished by observing that the dual polyhedron for the heterotic threefold is obtained
via a certain projection, which we explain, of the dual polyhedron of the fourfold.
This toric description is dual to the picture of the heterotic threefold in terms of a maximal
degeneration of the fourfold. We will argue that, at least for the models we consider here, the
relation between the fourfold and the heterotic threefold involves mirror symmetry in an interesting
way:
$$\eqalign{ 
\F[X] &= \Het[Z, V^X] \quad, \quad \widetilde{X} = (\widetilde{Z}, \cp1) \cropen{3pt}
\F[\widetilde{X}] &= \Het[\widetilde{Y}, V^{\widetilde{X}}] \quad, \quad X = (Y,\cp1)\cr}$$
where tildes are used to denote mirror manifolds and $X=(Y,\cp1)$, for example, denotes that $X$ is a
fibration by a (\cy) threefold $Y$ fibered over a $\cp1$.

Another important question is how to compute explicitly the $F$-theory superpotential on a given
fourfold and try to understand this as a superpotential for the corresponding heterotic theory. When
we do this several problems arise. At this point we are unable to give
a complete answer; we are presenting here various examples and related observations.

An immediate question that arises on computing a superpotential from the fourfold $X$ is that the 
superpotential is a cubic function of the K\"ahler parameters of $X$, while the heterotic
superpotential is a function of the volumes of the curves, and thus linear. The comparison makes
sense (so far) only for smooth divisors $D$ contributing to worldsheet instantons; curiously this
question has not arisen in examples that have been previously considered owing to the fact that in
these examples the cubic expression in the \K\ parameters is in fact trilinear. We discuss these
previous calculations in Section 7 (see also Section 5.2). It seems that the resolution is to regard
the volume of the elliptic fibre as an infinitessimal and in this limit the volumes of the divisors
are indeed trilinear.

Other questions arise when we compare the $F$-theory superpotential
corresponding to gauge groups, with a $d=3$ dimensional Yang-Mills theory, following\
\REFS{\KatzVafa}{S. Katz and C. Vafa, ``Geometric engineering of $N=1$ quantum
field theories'',\\ \npb{497} (1997) 196, hep-th/9611090.}
\REFSCON{\Vafatwo}{C. Vafa, ``On $N=1$ Yang-Mills in four dimensions'', hep-th/9801139.}
\refsend. 
Their work deals
with divisors corresponding to gauge groups~
\Ref{\BIKMSV}{M.~Bershadsky, K.~Intriligator, S.~Kachru, D.~R.~Morrison, V.~Sadov, and
C.~Vafa, ``Geometric singularities and enhanced gauge symmetries'',\\ 
\npb{481} (1996) 215, hep-th/9605200.}, 
arising as resolution of the Weierstrass model of the
elliptic fourfold: each (irreducible) divisor mapping via the elliptic fibration to the same surface
can be identified with a node in the extended Dynkin diagram of the group.
 A key point in their computation is that all (or none) of 
the irreducible divisors of a Dynkin diagram contribute to
the superpotential.  In Section 4 we consider  explicit examples where ``mixed
configurations" occur, that is when some divisors  contribute to the superpotential,
and some do not (following the criterion of~ 
\Ref{\Wittenone}{E. Witten, ``Nonperturbative Superpotentials in String Theory'',\\
\npb{474} (1996) 343, hep-th/9604030.}). 
This happens when the
gauge group is not simply laced ($SO(odd), \ Sp(n), \ G_2, \ F_4 $). A non simply laced group
arises via the action of monodromy on a group that is simply laced. In such a case the divisor
(two divisors for the case of $F_4$) that arise through the identification of divisors of the simply
laced group may have $\chi ({\cal O}_D) \neq 1$, violating the condition to contribute to the
superpotential. We shall refer to such divisors as being dissident. In { Section 6} we show that there
are many such examples. If $\chi ( {\cal O}_D) > 1 $,  $X$ turns out  not to be general in moduli
(in the toric case this correspond to the existence of
non-toric parameters) and that the `dissident' divisor $D$ will not remain an effective
divisor for general values of the complex moduli space. 
If $\chi ( {\cal O}_D) \leq 0$, then the divisor will be general in moduli and
we cannot reproduce Vafa's computations. On the other hand we show that in the toric
case $h^{2,1} (X) >0$: this always leads to an interesting structure on the heterotic
dual (see~
\REFS{\Diaconescu}{D. Diaconescu, private communication}
\REFSCON{\Ganor}{O. J. Ganor, ``A Note On Zeroes Of Superpotentials in F--Theory'',
\npb{499} 55 (1997), hep-th/9612077}
\refsend).

Many of our observations will be
recognisable to those who have developed a local description of the duality
in terms of branes wrapping the singular fibres of the fourfold seen as an elliptic
fibration over the threefold base $B^X$. On the other hand our observation is that
interesting features arise precisely as a result of trying to fit the local pieces
together into a global manifold. In particular there is a tendency for the gauge
group of the effective theory to `grow' since maintaining the fibration structure of
the fourfold $X$, when we put together the local singularities, requires
further resolution of these singularities.

The layout of this paper is the following: in \SS2 we gather together some
expressions that compute the Hodge numbers of a \cy\ fourfold $X$, corresponding to a
reflexive polyhedron $\nabla$,  as well as the arithmetic genus of its divisors in
terms of the combinatoric properties of $\nabla$. We find also interesting expressions, whose significance
we do not properly understand, that relate the arithmetic genera of the divisors of a manifold to the
arithmetic genera of the divisors of the mirror.  In \SS3 we introduce our basic model fourfold
$X$ and describe its structure. This model is perhaps as simple as one can have without taking a model that
is completely trivial. The gauge group that we have is $SU(2)\times G_2$. In \SS4 we explain the structure of
this model. We set out originally to construct a model with group $SU(2)\times SU(3)$ however
maintaining the fibration structure of our particular choice $X$ required extending the $SU(3)$ to
$G_2$ in a way that we explain in detail. We also show how to extend the gauge group by
taking the degenerate fibers to have a more complicated structure. The new element
here is the process by which we show how to divide the fans in such a way as to
maintain the fibration. In \SS5 we analyze the Yukawa coupling and the structure of the Mori
cone; for the Mori Cone we implement a procedure advocated in~
\Ref{\CoxKatz}{D. A. Cox and S. Katz, ``Mirror Symmetry and Algebraic Geometry'', AMS 1999.}. 
It is clear however that our implementation of this procedure leads to a cone that is too large.
This is similar to a recent result in~
\Ref{\Szendroi}{B. Szendr\"oi, ``On an Example of Aspinwall and Morrison'', math.AG/9911064.}.
In \SS7 we discuss the computation of the superpotential and compare this calculation to that for
previous~examples.
\newpage
\section{toric}{Toric Preliminaries}
We draw together here some essential results that we will need. 
\cy\ fourfolds, $X$, for which the holonomy is SU(4) rather than a subgroup have a Hodge
diamond whose top half is of the form
$$
\matrix{
&&&&&1&&&&\cr
&&&&0&&0&&&\cr
&&&0&&h^{11}&&0&&\cr
&&0&&h^{21}&&h^{21}&&0&\cr
&1~&&h^{31}&&h^{22}&&h^{31}&&~1~.\cr
}$$
\vskip5pt
There is also a linear relation between the Hodge numbers
$$
h^{22}~=~2\,(22 + 2h^{11} + 2h^{31} - h^{21})~.$$
Thus there are three independent Hodge numbers and the Euler number 
is given in terms of the Hodge numbers by
$$
\ch_E(X)~=~6\,(8 + h^{11} + h^{31} - h^{21})~.\eqlabel{chiphys}$$

The three independent Hodge numbers $h^{31}$, $h^{21}$ and $h^{11}$ may be
determined directly from the Newton polyhedron, $\D$, and its dual, $\nabla$, (our
convention is that the fan of $X$ is that fan over the faces of $\nabla$) via the
expressions~ 
\REFS\Batyrev{V. Batyrev, J.\ Alg.\ Geom. {\bf 3} (1994) 493--535,
alg-geom/9310003.}
\REFSCON\Aspinwalletal{P. S. Aspinwall, B. R. Greene and D. R. Morrison,
``\cy\ Moduli Space, Mirror Manifolds and Space-Time Topology Change in 
String Theory'', \\
\npb{416} (1994) 414, hep-th/9309097.} 
\REFSCON{\KLRY}{A. Klemm, B. Lian, S. S. Roan and S. T. Yau, 
``\cy\ Fourfolds for $M$ Theory and $F$ Theory Compactifications'', 
\npb{518} (1998) 515, hep-th/9701023.} 
\refsend
$$\eqalign{
h^{31} &= \pts(\D) - \hskip-10pt\sum_{\codim\,\th = 1}\hskip-10pt\inte(\th) + 
\hskip-10pt\sum_{\codim\,\th = 2}\hskip-10pt\inte(\th)\,\inte(\tilde\th) -
6\cropen{10pt}
h^{11} &= 
\pts(\nabla) - \hskip-10pt\sum_{\codim\,\tilde\th = 1}\hskip-10pt\inte(\tilde\th) + 
\hskip-10pt\sum_{\codim\,\tilde\th = 2}\hskip-10pt\inte(\tilde\th)\,\inte(\th) -
6\cropen{10pt}
h^{21} &= 
\hskip-10pt\sum_{\codim\,\th = 3}\hskip-10pt\inte(\tilde\th)\,\inte(\th)\cr}
\eqlabel{hnos} $$ 
In these expressions $\pts(\D)$
denotes the number of lattice points in $\D$, while $\th$ runs over the faces of $\D$,
$\inte(\th)$ denotes the number of lattice points  strictly interior to $\th$ and
$\tilde\th$ denotes the face of $\nabla$ dual to $\th$.
It is interesting to note that Batyrev
\Ref{\Batyrevtwo}{D. Dais and V. Batyrev,\\
``Strong McKay Correspondence, String-Theoretic Hodge Numbers 
and Mirror Symmetry'', Topology 35, No.4, 901 (1996), alg-geom/9410001.}\ 
gives also an expression for the ``physicist's
Euler number'' in terms of the {\sl volumes\/} of the faces of the polyhedra
$$
\ch'_E~=~\hskip-10pt\sum_{\codim\,\th = 2}\hskip-10pt\vol(\th)\,\vol(\tilde\th) -
2\hskip-10pt\sum_{\codim\,\th = 3}\hskip-10pt\vol(\th)\,\vol(\tilde\th) +
\hskip-10pt\sum_{\codim\,\th = 4}\hskip-10pt\vol(\th)\,\vol(\tilde\th)
\eqlabel{chiBat}$$  
in which the volumes are
normalized such that the volume of a fundamental lattice simplex is unity (rather
than $1/d!$ in $d$ dimensions). If the fourfold is nonsingular then these two
expressions for the Euler number yield the same result. There are cases however of
fourfolds $X_{mnp}$ defined by \eqref{charges} for which
$\ch'_E\neq\ch_E$ indicating that these fourfolds are singular (for all values of
their parameters). 

Each point $q\in\nabla$ that is not interior to a facet (codimension one face)
corresponds to a divisor $D_q$ of $X$. Such a point $q$ defines a face
$\tilde\th_q$ of $\nabla$, the unique face to which it is interior\Footnote{A point
$q\in\nabla$ that is not the interior point is contained in some face of $\nabla$. We ask
whether $q$ is interior to this face or if it lies in the boundary. If it lies in the boundary
then it lies in a face of lower dimension and we ask again if it lies in the interior to this
face or in the boundary. Proceeding in this way we either arrive at the unique face to which
$q$ is interior or we find that $q$ is a vertex. It is therefore convenient to adopt the
convention that the vertices are interior to themselves.\smallskip}, and in virtue of
duality also a face $\th_q$ of $\D$. Klemm {\it et al.\/}~\cite{\KLRY}
establish an elegant expression for the arithmetic genus, $\ch_q$, of $D_q$:
$$
\ch_q~=~\ch\left(\ca{O}(D_q)\right)~=~
1 - (-1)^{\dim(\tilde\th_q)}\inte(\th_q)~.\eqlabel{divisors}$$

An observation, made already in \cite{\KLRY}, is that manifolds containing divisors of
arithmetic genus one are abundant. These  exist whenever $\nabla$ has faces that (a)
have interior points and that (b) are dual to faces of $\D$ that have no interior
points. Klemm {\it et al.\/} study and interesting class of manifolds that we will term
the KLRY spaces, and whose construction we review in Sect.~\chapref{X}.  All these
manifold  and all their mirrors have such faces\Footnote{The number of these divisors is
by
\eqref{divisors} always finite. The infinite number found in~ 
\Ref{\DGW}{R. Donagi, A. Grassi and E. Witten,
``A non-perturbative superpotential with $E_8$  symmetry'',
Mod. Phys. Lett. {\bf A}17 (1996), hep-th/9607091.}\ 
is  associated with the fact the manifold under consideration is not in the class that we
consider here, since it cannot be realised as a hypersurface in a toric variety given by
a single equation, and hence by a single polyhedron.  It would be of interest to study
this example from the toric perspective.  The formalism however is much less developed for
the cases that require more than one equation.}.

Witten~\cite{\Wittenone} shows that $\ch(\ca{O}(D_q) )=1$ 
is a necessary condition to contribute to the superpotential, while
 $$
h^0(\ca{O}(D_q))=1~,~~~ h^i(\ca{O}(D_q))=0~, \ i>0$$ 
is a sufficient condition.

It is of interest to note that in the expressions \eqref{hnos} for the Hodge numbers
there occur crossterms that involve both $\nabla$ and $\D$
\smallskip
 $$\eqalign{
\bar\d~&=~\hskip-10pt\sum_{\codim\,\tilde\th = 2}\hskip-10pt
\inte(\th)\,\inte(\tilde\th)~=~\sum_{q\in(\dim\tilde\th=3)}(\ch_q -1)\cropen{10pt}
h^{21}~&=~\hskip-10pt\sum_{\codim\,\tilde\th = 3}\hskip-10pt
\inte(\th)\,\inte(\tilde\th)
~=\hskip3pt-\hskip-10pt\sum_{q\in(\dim\tilde\th=2)}(\ch_q-1)\cropen{10pt}
\d~&=~\hskip-10pt\sum_{\codim\,\tilde\th = 4}\hskip-10pt
\inte(\th)\,\inte(\tilde\th)~=~\sum_{q\in(\dim\tilde\th=1)}(\ch_q -1)\cropen{10pt}\cr}
\eqlabel{crossterms}$$
\vskip-10pt 
The quantities $\d$ and $\bar\d$ are respectively
the number of non-toric deformations of
$X$ and its mirror, while $2h^{21}$ is the number of cohomology classes of three cycles. Since
the KLRY formula\eqref{divisors} shows that all $q$'s interior to a given $\tilde\th_q$ have
the same arithmetic genus it follows that these crossterms are also equal to $\pm(\ch_q-1)$ as
given. If $\hbox{dim}(\tilde\th)=4$ then $\th_q$ is a vertex and $\inte(\th_q)=1$ by convention
so
$$\sum_{\codim(\tilde\th)=1}\hskip-10pt(\ch_q-1)=
 \hskip5pt-\hskip-5pt\sum_{\codim(\tilde\th)=1}\hskip-10pt\inte(\tilde\th_q)$$
which counts the divisors of $\IP_\nabla$ that do not intersect the hypersurface. A dual
relation obtains also for $\hbox{dim}(\tilde\th)=0$. We summarize these relations in the following
table:
\smallskip 
$$
\def\skip{\hskip7pt}
\vbox{\offinterlineskip\tabskip=0pt\halign{%
\vrule height12pt depth 8pt%
\hfil\skip$#$\skip\hfil\vrule&\hfil\skip$#$\skip\hfil\vrule&\hfil\skip$#$\skip\hfil\vrule\cr
\noalign{\hrule}
\hbox{dim}(\tilde\th_q)&\hbox{dim}(\th_q)&\hbox{$(\ch_q-1)$ contributes to}\cr
\noalign{\hrule\vskip3pt\hrule}
\omit\vrule height2pt\hfil\vrule &&\cr
4&0&\hbox{$q$'s that are not divisors}\cr
3&1&\tilde\d\cr
2&2&~~h^{21}\cr
1&3&\d\cr
0&4&\hbox{irrelevant monomials}\cr
\omit\vrule height2pt\hfil\vrule &&\cr
\noalign{\hrule}
}}$$
\smallskip\noindent
Note that if a divisor has $\ch\neq 1$ then this divisor contributes to precisely one of the
quantities in the third column of the table.

There are also curious relations whose significance we do not understand such as
 $$
\sum_{q\in\tilde\th}(\ch_q-1)~=~\sum_{\tilde{q}\in\th}(\ch_{\tilde{q}}-1)$$
that relate $\ch-1$ for divisors in the manifold and its mirror. Another such relation is
 $$
\sum_{q\in\partial\nabla}(\ch_q-1)~=~
{\chi_E(X)\over 6} - \bigl(\pts(\nabla)+\pts(\D)\bigr) + 4$$
or equivalently
 $$
\sum_{q\in\partial\nabla}\ch_q~=~
{\chi_E(X)\over 6} - \pts(\D) + 3$$ 
where $\partial\nabla$ denotes $\nabla$ less the interior point and owing to our convention
$\ch_q-1=-1$ for $q$ in a codimension one face of $\nabla$.
\newpage
\section{heterotic}{Heterotic Structure}
\vskip-30pt
\subsection{Fibrations}
A first statement of the relation between the manifolds that appear in \eqref{duals} 
may be made in terms of fibrations.  A poor but useful notation that we employ is to
write $(\cal F, \cal B)$  for a manifold that is a fibration over a base $\cal B$ 
with generic fiber $\cal F$.  The notation is poor since a manifold is not uniquely
specified.  There may well be different manifolds that can be realized as
fibrations over a given base with the same generic fiber, the difference being due to
the manner in which the fibers degenerate over subvarieties of the base. This said,
the relation between the manifolds of \eqref{duals} is believed to be the following:
$$\eqalign{ 
X = ({\cal E}, B^X) = (K3^Y, &B^Z)\quad ,\quad Z =({\cal E}, B^Z)\cr
B^X = (\cp1, B^Z)\quad ,&\quad K3^Y = (\ca{E}, \cp1)\cr}\eqlabel{manifolds}$$
with $\cal E$ denoting an elliptic curve and $K3^Y$ denoting a $K3$ manifold. The
superfix $Y$ refers to another \cy\ threefold to which we shall refer as we proceed. 
In other words, $X$ is an elliptic  fibration over a base $B^X$, with $B^X$ a
$\cp1$-fibration over a two dimensional base $B^Z$. The manifold $Z$ of the heterotic
compactification is then an elliptic fibration over this same two dimensional base.
The second representation of $X$ states that it is also a fibration over $B^Z$ with
fiber an elliptic~K3.

The purpose of the present paper is, in part, to study the relations between these 
manifolds in the toric context for which some degree of control is afforded by the
relation between \cys\ and reflexive polyhedra and the observation of~ 
\Ref{\Avrametal}{A. Avram, M. Mandelberg, M. Kreuzer and H. Skarke,\\ 
``Searching for $K3$ Fibrations'', \npb{505} (1997) 625, hep-th/9610154.}\ 
that the fibration structure of
a manifold specified by such a polyhedron is visible in the polyhedron.  Although the
nature of the bundle
$V^X$ is not properly understood in terms of the toric data, nevertheless some 
information is available. For example, the structure group $G^X$ of $V^X$ can be read
off from the polyhedron~
\REFS\CandelasFont{P.~Candelas and A.~Font, ``Duality Between the Webs of 
Heterotic and Type II Vacua'', \npb{511} (1998) 295, hep-th/9603170.} 
\REFSCON{\CPR}{P. Candelas, E. Perevalov and G. Rajesh,\\ 
``Toric Geometry and Enhanced Gauge Symmetry of F Theory/Heterotic Vacua'',\\
\npb{507} (1997) 445, hep-th/9704097}
\refsend.  
Another issue that we study in the toric context
is that of the existence of divisors that give rise to a superpotential through
non-perturbative effects.  Proceeding loosely, the integral points, $q$, of the dual
polyhedron, $\nabla^X$, of $X$ are in direct correspondence with the divisors of~$X$. 
Moreover, Klemm {\it et al.\/}\cite{\KLRY} have formulated a simple and elegant criterion
that distinguishes the points corresponding to the divisors of arithmetic genus one. The
toric context, though not general, permits a study of examples and a certain
systematization which we find useful.

As noted above a reflexive polyhedron corresponding to a \cym\ that is a fibration
$(\ca{F}, \ca{B})$ has a slice corresponding to the dual polyhedron of the fiber
$\ca{F}$ and this enables us to establish a standard coordinate system for the
polyhedra. It is perhaps easiest to see this at work in an example. The first column
of Table~\tabref{nablas} lists the integral points of the dual polyhedron of an
interesting example which we shall denote by $X$ throughout this article.
\pageinsert
\def\nablaX{%
\def\extraheight{\multispan 7\vrule height 2pt\hfil\vrule\cr}
\vbox{\offinterlineskip\tabskip=0pt\halign{%
\vrule height9pt depth 4pt\hfil\hskip4pt
(##&\hfil##,&\hfil##,&\hfil##,&\hfil##,&\hfil##&##)\hskip4pt\hfil\vrule\cr
\noalign{\hrule}
\multispan 7\vrule height12pt
depth 6pt\hfil$\mathstrut\hskip5pt\nabla^X$\hfil\vrule\cr
\noalign{\hrule}
\extraheight
&$x_1$&$x_2$&$x_3$&$x_4$&$x_5$&\cr
\extraheight
\noalign{\hrule\vskip3pt\hrule}
\extraheight
&-1&0&0&2&3&\cr
&0&-1&0&2&3&\cr
&0&0&-1&2&3&\cr
&0&0&-1&1&2&\cr
&0&0&0&-1&0&\cr
&0&0&0&0&-1&\cr
&0&0&0&0&0&\cr
&0&0&0&0&1&\cr
&0&0&0&1&1&\cr
&0&0&0&1&2&\cr
&0&0&0&2&3&\cr
&0&0&1&1&2&\cr
&0&0&1&2&3&\cr
&0&0&2&2&3&\cr
&0&0&1&1&1&\cr
&0&1&2&2&3&\cr
&0&1&3&2&3&\cr
&1&0&4&2&3&\cr
\extraheight
\noalign{\hrule}
}}}
\def\nablaY{%
\def\extraheight{\multispan 6\vrule height 2pt\hfil\vrule\cr}
\vbox{\offinterlineskip\tabskip=0pt\halign{%
\vrule height9pt depth 4pt\hfil\hskip15pt%
(##&\hfil##,&\hfil##,&\hfil##,&\hfil##&##)\hskip4pt\hfil\vrule\cr
\noalign{\hrule} 
\multispan 6\vrule height12pt depth 6pt\hfil
$\mathstrut\hskip5pt\nabla^Y$\hfil\vrule\cr
\noalign{\hrule}
\extraheight
\multispan 6\vrule height10pt depth
5pt\hskip4pt$(\;0,x_2,x_3,x_4,x_5)$\hskip4pt\vrule\cr
\extraheight
\noalign{\hrule\vskip3pt\hrule}
\extraheight
&-1&0&2&3&\cr
&0&-1&2&3&\cr
&0&-1&1&2&\cr
&0&0&-1&0&\cr
&0&0&0&-1&\cr
&0&0&0&0&\cr
&0&0&0&1&\cr
&0&0&1&1&\cr
&0&0&1&2&\cr
&0&0&2&3&\cr
&0&1&1&2&\cr
&0&1&2&3&\cr
&0&2&2&3&\cr
&0&1&1&1&\cr
&1&2&2&3&\cr
&1&3&2&3&\cr
\extraheight
\noalign{\hrule} 
}}}
\def\nablaZ{%
\def\extraheight{\multispan 6\vrule height 2pt\hfil\vrule\cr}
\vbox{\offinterlineskip\tabskip=0pt\halign{%
\vrule height9pt depth 4pt\hfil\hskip15pt%
(##&\hfil##,&\hfil##,&\hfil##,&\hfil##&##)\hskip4pt\hfil\vrule\cr
\noalign{\hrule} 
\multispan 6\vrule height12pt depth 6pt\hfil
$\mathstrut\hskip5pt\nabla^Z$\hfil\vrule\cr
\noalign{\hrule}
\extraheight
\multispan 6\vrule height10pt depth
5pt\hskip3pt$(x_1,x_2,\widehat{x_3},x_4,x_5)$\hskip3pt\vrule\cr
\extraheight
\noalign{\hrule\vskip3pt\hrule}
\extraheight
&-1&0&2&3&\cr 
&0&-1&2&3&\cr 
&0&0&-1&0&\cr 
&0&0&0&-1&\cr 
&0&0&0&0&\cr 
&0&0&0&1&\cr 
&0&0&1&1&\cr 
&0&0&1&2&\cr
&0&0&2&3&\cr 
&0&1&2&3&\cr 
&1&0&2&3&\cr
\extraheight
\noalign{\hrule} 
}}}
\def\nablaKY{%
\def\extraheight{\multispan 5\vrule height 2pt\hfil\vrule\cr}
\vbox{\offinterlineskip\tabskip=0pt\halign{%
\vrule height9pt depth 4pt\hfil\hskip21pt%
(##&\hfil##,&\hfil##,&\hfil##&##)\hskip4pt\hfil\vrule\cr
\noalign{\hrule} 
\multispan 5\vrule height14pt depth 6pt\hfil
$\mathstrut\hskip5pt\nabla^{K3^Y}$\hfil\vrule\cr
\noalign{\hrule}
\extraheight
\multispan 5\vrule height10pt depth
5pt\hskip4pt$(\;0,\;0,x_3,x_4,x_5)$\hskip4pt\vrule\cr
\extraheight
\noalign{\hrule\vskip3pt\hrule}
\extraheight
&-1&2&3&\cr
&-1&1&2&\cr
&0&-1&0&\cr
&0&0&-1&\cr
&0&0&0&\cr
&0&0&1&\cr
&0&1&1&\cr
&0&1&2&\cr
&0&2&3&\cr
&1&1&2&\cr
&1&2&3&\cr
&2&2&3&\cr
&1&1&1&\cr
\extraheight
\noalign{\hrule} 
}}}
\def\nablaKZ{%
\def\extraheight{\multispan 5\vrule height 2pt\hfil\vrule\cr}
\vbox{\offinterlineskip\tabskip=0pt\halign{%
\vrule height9pt depth 4pt\hfil\hskip21pt%
(##&\hfil##,&\hfil##,&\hfil##&##)\hskip4pt\hfil\vrule\cr
\noalign{\hrule} 
\multispan 5\vrule height14pt depth 6pt\hfil
$\mathstrut\hskip5pt\nabla^{K3^Z}$\hfil\vrule\cr
\noalign{\hrule}
\extraheight
\multispan 5\vrule height10pt depth
5pt\hskip2pt$(x_1,\widehat{x_2},\widehat{x_3},x_4,x_5)$\hskip2pt\vrule\cr
\extraheight
\noalign{\hrule\vskip3pt\hrule}
\extraheight 
&-1&2&3&\cr 
&0&-1&0&\cr 
&0&0&-1&\cr 
&0&0&0&\cr 
&0&0&1&\cr 
&0&1&1&\cr 
&0&1&2&\cr
&0&2&3&\cr 
&1&2&3&\cr 
\extraheight
\noalign{\hrule} 
}}}
\def\nablaE{%
\def\extraheight{\multispan 4\vrule height 2pt\hfil\vrule\cr}
\vbox{\offinterlineskip\tabskip=0pt\halign{%
\vrule height9pt depth 4pt\hfil\hskip27pt%
(##&\hfil##,&\hfil##&##)\hskip4pt\hfil\vrule\cr
\noalign{\hrule} 
\multispan 4\vrule height12pt depth 6pt\hfil
$\mathstrut\hskip5pt\nabla^\ca{E}$\hfil\vrule\cr
\noalign{\hrule}
\extraheight
\multispan 4\vrule height10pt depth
5pt\hskip4pt$(\;0,\;0,\;0,x_4,x_5)$\hskip4pt\vrule\cr
\multispan 4\vrule height10pt depth
5pt\hskip4pt$(\widehat{x_1},\widehat{x_2},\widehat{x_3},x_4,x_5)$\hskip4pt\vrule\cr
\extraheight
\noalign{\hrule\vskip3pt\hrule}
\extraheight
&-1&0&\cr
&0&-1&\cr
&0&0&\cr
&0&1&\cr
&1&1&\cr
&1&2&\cr
&2&3&\cr
\extraheight
\noalign{\hrule}
}}}
\font\bigcal cmsy10 at 17pt
\def\hvbox#1{\hbox{\vbox to 7.7truein{\vfil #1\vfil}}}
\def\symbox#1{\hbox to 20pt{\hfil\bigcal\char #1\hfil}}
\centerline{
\hvbox{\vskip70pt\nablaX}
\hvbox{\vskip2.5truein\symbox{'056}\vskip2.5truein\symbox{'046}\vfill}
\hvbox{\nablaY\vskip20pt\nablaZ}
\hvbox{\vskip1.8truein\symbox{'040}\vskip4.0truein\symbox{'041}\vfill}
\hvbox{\vskip29pt\nablaKY\vskip20pt\nablaKZ\vskip15pt}
\hvbox{\vskip2.6truein\symbox{'055}\vskip2.6truein\symbox{'045}\vfill}
\hvbox{\vskip80pt\nablaE}
}
\vskip20pt\leavevmode\hskip-20pt
\tablecaption{nablas}{The dual polyhedron for $X$ with the data of the associated
fibrations. The $\nabla$'s on the upper level are linked by a series of injections
while those on the lower level are related by projections.}
\vfil
\endinsert
We take the Cartesian coordinates $(x_1,\,x_2,\,x_3,\,x_4,\,x_5)$ for the $\IR^5$ in
which the polyhedron is embedded. The points that lie in the hyperplane $\{x_1 = 0\}$
form
$\nabla^Y$, the dual polyhedron of a \cy\ threefold, $Y$. These points are listed in
the upper table of column two of Table~\tabref{nablas}. The points with $\{x_1 = x_2 =
0\}$ form
$\nabla^{K3^Y}$, the dual polyhedron of a $K3$ surface associated to $Y$. The points
with 
$\{x_1 = x_2 = x_3 = 0\}$ form the dual polyhedron, $\nabla^{\ca{E}}$, of the
Weierstrass torus
$\ca{E}=\IP^{(1,2,3)}[6]$. 
Finally the three points with $\{x_1 = x_2 = x_3 = x_4 = 0\}$ are the dual polyhedron
of a zero-dimensional \cym\Footnote{We can think of a zero-dimensional \cym\ as
$\IP_1[2]$ with equation \hbox{$\x_1^2 +\x_1\x_2 +\x_2^2 = 0$}. The Newton polyhedron
associated with this equation consists of three points in a straight line. After a
change of coordinates these become the points $x=-1,0,1$. This trivial reflexive
polyhedron is self-dual}. The lower route through the table is realised by making the
indicated projections. The coordinates with hats are projected out in this process.
Thus $\nabla^X\to\nabla^Z$ corresponds to the projection 
$(x_1,x_2,x_3,x_4,x_5)\to (x_1,x_2,x_4,x_5)$. Note that the fibration manifests
additional structure, not only does the dual polyhedron of ${\cal E}$ appear as a
slice but also as a projection onto the last two coordinates. This is related to the
fact that $\nabla^\ca{E}$ is self dual.

We would like now to discuss the bases of the elliptic \cy\ and $K3$ fibrations which
we denote by $B^X$, $B^Y$ and $B^Z$ respectively. We have noted that we can see the
dual polyhedron of the fiber, $\ca{E}$, of the fibration $X = ({\cal E}, B^X)$ as the
slice 
$\{x_1 = x_2 = x_3 = 0 \}$ of $\nabla^X$.  The base
$B^X$ may also be seen as the {\sl projection} onto the first three coordinates.  This
gives us the points of the first column of Table~\tabref{bases} the rays from the
origin through these points yield the {\sl fan} of
$B^X$.   Note also that $B^X$ can be obtained not only as the projection
to the first three coordinates but also as the slice $\{ x_4 = 2, x_5 = 3\}$, which
is a three-face of $\nabla^X$.  It is one of the observations of \cite{\Avrametal} that
the roles of injections and projections are interchanged by mirror symmetry so the fact
that $\nabla^{\cal E}$ and
$B^X$ are visible both as injections and projections has the consequence that the
mirror, $\widetilde X$, of $X$ is also an elliptic fibration 
$\widetilde X = ({\cal E}, B^{\widetilde X})$, where in this relation 
$B^{\widetilde X}$ denotes the base of the mirror fibration and we write ${\cal E}$ 
for the fiber in place of
$\widetilde{\cal E}$ since $\nabla^{\cal E}$ is self mirror.  $B^X$ is not \cy\ so
there is no notion of a mirror of $B^X$. For the \cy\ fibration 
$X = (Y,\cp1)$ we have already noted that $\nabla^Y$ is the slice $\{x_1 = 0\}$ of
$\nabla^X$. The fan of the \cp1\ consists of the three points $x_1 = -1,0,1$ obtained
by projecting $\nabla^X$ onto the first coordinate. The threefold $Y$ is itself an
elliptic fibration, $Y=(\ca{E},B^Y)$, over a base $B^Y$ whose toric data are obtained
by projecting $\nabla^Y$ to its first two coordinates. This gives the upper table of
the second column of Table~\tabref{bases}. 
As for the $K3$-fibration $X = (K3^Y, B^Z)$, we have seen the $K3^Y$ as the
slice $\{ x_1 = x_2 = 0\}$ and we see $B^Z$ as the
projection onto the first two coordinates, the result being the lower table of
the second column of Table~\tabref{bases}.  The base of the $K3$-fibration is
in fact the base of the fibration $Z = ({\cal E}, B^Z)$ as we shall see.
Note however that $Y$ and the $K3$ are not projections onto any slices
so we might not expect the mirror of $X$ to be a \cy\ and $K3$-fibration, although
we shall see presently that it is.  
\midinsert
\def\BX{%
\def\extraheight{\multispan 5\vrule height 2pt\hfil\vrule\cr}
\vbox{\offinterlineskip\tabskip=0pt\halign{%
\vrule height9pt depth 4pt\hfil\hskip21pt%
(##&\hfil##,&\hfil##,&\hfil##&##)\hskip4pt\hfil\vrule\cr
\noalign{\hrule} 
\multispan 5\vrule height14pt depth 6pt\hfil
$\mathstrut\hskip5pt B^X$\hfil\vrule\cr
\noalign{\hrule}
\extraheight
\multispan 5\vrule height10pt depth
5pt\hskip4pt$(x_1,x_2,x_3,\widehat{x_4},\widehat{x_5})$\hskip4pt\vrule\cr
\multispan 5\vrule height10pt depth
5pt\hskip4pt$(x_1,x_2,x_3,\;2,\;3)$\hskip4pt\vrule\cr
\extraheight
\noalign{\hrule\vskip3pt\hrule}
\extraheight
&-1&0&0&\cr
&0&-1&0&\cr
&0&0&-1&\cr
&0&0&0&\cr
&0&0&1&\cr
&0&0&2&\cr
&0&1&2&\cr
&0&1&3&\cr
&1&0&4&\cr
\extraheight
\noalign{\hrule}
}}}
\def\BY{%
\def\extraheight{\multispan 4\vrule height 2pt\hfil\vrule\cr}
\vbox{\offinterlineskip\tabskip=0pt\halign{%
\vrule height9pt depth 4pt\hfil\hskip27pt%
(##&\hfil##,&\hfil##&##)\hskip4pt\hfil\vrule\cr
\noalign{\hrule} 
\multispan 4\vrule height12pt depth 6pt\hfil
$\mathstrut\hskip5pt B^Y$\hfil\vrule\cr
\noalign{\hrule}
\extraheight
\multispan 4\vrule height10pt depth
5pt\hskip4pt$(\;0,x_2,x_3,\widehat{x_4},\widehat{x_5})$\hskip4pt\vrule\cr
\multispan 4\vrule height10pt depth
5pt\hskip4pt$(\;0,x_2,x_3,\;2,\;3)$\hskip4pt\vrule\cr
\extraheight
\noalign{\hrule\vskip3pt\hrule}
\extraheight
&-1&0&\cr
&0&-1&\cr
&0&0&\cr
&0&1&\cr
&0&2&\cr
&1&2&\cr
&1&3&\cr
\extraheight
\noalign{\hrule}
}}}
\def\BZ{%
\def\extraheight{\multispan 4\vrule height 2pt\hfil\vrule\cr}
\vbox{\offinterlineskip\tabskip=0pt\halign{%
\vrule height9pt depth 4pt\hfil\hskip27pt%
(##&\hfil##,&\hfil##&##)\hskip4pt\hfil\vrule\cr
\noalign{\hrule} 
\multispan 4\vrule height12pt depth 6pt\hfil
$\mathstrut\hskip5pt B^Z$\hfil\vrule\cr
\noalign{\hrule}
\extraheight
\multispan 4\vrule height10pt depth
5pt\hskip4pt$(x_1,x_2,\widehat{x_3},\widehat{x_4},\widehat{x_5})$\hskip4pt\vrule\cr
\multispan 4\vrule height10pt depth
5pt\hskip4pt$(x_1,x_2,\widehat{x_3},\;2,\;3)$\hskip4pt\vrule\cr
\extraheight
\noalign{\hrule\vskip3pt\hrule}
\extraheight
&-1&0&\cr 
&0&-1&\cr 
&0&0&\cr 
&0&1&\cr 
&1&0&\cr 
\extraheight
\noalign{\hrule}
}}}
\def\PZ{%
\def\extraheight{\multispan 3\vrule height 2pt\hfil\vrule\cr}
\vbox{\offinterlineskip\tabskip=0pt\halign{%
\vrule height9pt depth 4pt\hfil\hskip33pt%
(##&\hfil##&##)\hskip4pt\hfil\vrule\cr
\noalign{\hrule} 
\multispan 3\vrule height12pt depth 6pt\hfil
$\mathstrut\hskip5pt \cp1^Z$\hfil\vrule\cr
\noalign{\hrule}
\extraheight
\multispan 3\vrule height10pt depth
5pt\hskip4pt$(0\;,x_2,\widehat{x_3},\widehat{x_4},\widehat{x_5})$\hskip4pt\vrule\cr
\multispan 3\vrule height10pt depth
5pt\hskip4pt$(0\;,x_2,\widehat{x_3},\;2,\;3)$\hskip4pt\vrule\cr
\extraheight
\noalign{\hrule\vskip3pt\hrule}
\extraheight
&-1&\cr 
&0&\cr 
&1&\cr 
\extraheight
\noalign{\hrule}
}}}
\def\PY{%
\def\extraheight{\multispan 3\vrule height 2pt\hfil\vrule\cr}
\vbox{\offinterlineskip\tabskip=0pt\halign{%
\vrule height9pt depth 4pt\hfil\hskip33pt%
(##&\hfil##&##)\hskip4pt\hfil\vrule\cr
\noalign{\hrule} 
\multispan 3\vrule height12pt depth 6pt\hfil
$\mathstrut\hskip5pt \cp1^Y$\hfil\vrule\cr
\noalign{\hrule}
\extraheight
\multispan 3\vrule height10pt depth
5pt\hskip2.5pt$(x_1,\widehat{x_2},\widehat{x_3},\widehat{x_4},\widehat{x_5})$%
\hskip2.5pt\vrule\cr
\multispan 3\vrule height10pt depth
5pt\hskip2pt$(x_1,\widehat{x_2},\widehat{x_3},\;2,\;3)$\hskip2pt\vrule\cr
\extraheight
\noalign{\hrule\vskip3pt\hrule}
\extraheight
&-1&\cr 
&0&\cr 
&1&\cr 
\extraheight
\noalign{\hrule}
}}}
\font\bigcal cmsy10 at 17pt
\def\hvbox#1{\hbox{\vbox to 5truein{\vfil #1\vfil}}}
\def\symbox#1{\hbox to 20pt{\hfil\bigcal\char #1\hfil}}
\centerline{
\hvbox{\BX}
\hvbox{\vskip2truein\symbox{'056}\vskip1.5truein\symbox{'046}\vfill}
\hvbox{\BY\vskip20pt\BZ}
\hvbox{\vskip1.8truein\symbox{'040}\vskip2.1truein\symbox{'041}\vfill}
\hvbox{\vskip25pt\PZ\vskip20pt\PY}
}
\vskip20pt\leavevmode\hskip-20pt
\tablecaption{bases}{The points corresponding to the bases of the fibrations.}
\vfil
\endinsert
To summarize thus far, the coordinates of $\nabla^X$ relate $B^X$, $B^Y$, $B^Z$ and
\ca{E}\ as follows  
 $$\eqalign{
&\hskip25pt B^X\hskip27pt\ca{E}\cropen{-7pt}
&\hphantom{~(}\overbrace{\hphantom{x_1,\,x_2,\,x_3}}\,\,
\overbrace{\hphantom{x_4,\,x_5}}\cropen{-10pt}
\vec{x}~=
&~(\underbrace{\strut x_1,\,x_2},\,x_3,\,x_4,\,x_5)\cropen{-3pt} 
&\hskip15pt B^Z\cropen{-20pt}
&\hphantom{~(x_1,\,}\underbrace{\strut \hphantom{x_2,\,x_3}}\cropen{-3pt} 
&\hskip31pt B^Y\cr}$$

The reader wishing to acquire dexterity with seeing the various fibrations should
check from Table~\tabref{bases} that
 $$
B^X~=~(\cp1,\, B^Z)~~~~\hbox{and}~~~~B^X~=~(B^Y,\,\cp1)~.$$
 We want to find $Z=(\ca{E}, B^Z)$ in $\nabla^X$ so we need
to project out the coordinate $x_3$. Thus
$\nabla^Z$ can be realized by performing the projection
$(x_1,x_2,x_3,x_4,x_5)\mapsto (x_1,x_2,x_4,x_5)$. 
This yields the sixth column of Table~\tabref{nablas}. Another way of stating this is
that $Z$ is being identified as the base of a fibration. This statement seems to
involve mirror symmetry in a nontrivial way. The fibrations being
 $$\eqalign{
X~&=~(Y,\,\cp1)~=~\left(\IZ_2,\,\IP(\nabla^Z)\right)\cr
\widetilde{X}~&=~(\widetilde{Z},\,\cp1)~=~\left(\IZ_2,\,\IP(\Delta^Y)\right)\cr}$$
where $\IP(\nabla^Z)$ and $\IP(\Delta^Y)$ denote the toric manifolds corresponding to
the fans over the faces of $\nabla^Z$ and $\Delta^Y$ respectively.

A class of fourfolds $X$ with the structure $X = ({\cal E}, B^X)$ with 
\hbox{$B^X = (\cp1, B^Z)$} has been discussed in \cite{\KLRY}.  The idea is to take
$B^X$ of the form $(\cp1, (\cp1, \cp1))$.  The inner fibration $(\cp1, \cp1)$ is
taken to be a fiber bundle, the Hirzebruch surface\Footnote{Notice that this already
shows how loose the notation $(\cp1, \cp1)$ is since even for the case of a fiber
bundle there is one of these for each integer $m$.} $\IF_m$.  The wrapping of the
outer $\cp1$ is specified by two further integers corresponding to the wrapping of
this
$\cp1$ about each of the other two.  The resulting manifold is denoted by $\IF_{mnp}$ 
in \cite{\KLRY}.  For many values of these integers, it is possible to define in terms of 
toric data manifolds $ X = ({\cal E}, \IF_{mnp})$ in such a way that $X$ is \cy.  By
this we mean here that we can associate a Newton polyhedron with $X$ and this
polyhedron has the property of being reflexive.

These reflexive polyhedra have interesting structure corresponding to the fact that 
in addition to the structure \eqref{duals} we have also
$$X = (Y,\cp1) \quad, \quad Y = (K3^Y, \cp1) \quad, \quad K3^Y = ({\cal E}, \cp1)
     \quad, \quad {\cal E} = (\IZ_2 , \cp1) \quad ,\eqlabel{Xfibrations}$$
with $Y$ a \cy\ threefold.  Now, this structure manifests itself in the dual 
polyhedron $\nabla^X$ of $X$ in a very simple way: $\nabla^X$ contains a codimension
one slice which is $\nabla^Y$, the dual polyhedron of $Y$ and this structure is
repeated; $\nabla^Y$ contains $\nabla^{K3^Y}$, the dual polyhedron of the $K3$ surface
as a slice and $\nabla^{K3^Y}$ contains $\nabla^{\cal E}$ the triangle of the
Weierstrass polynomial as a slice. Finally, $\nabla^{\cal E}$ contains a line with
three points corresponding to a zero dimensional \cym.  In other words we have a
series of injections
$$
\nabla^X\lla\nabla^Y\lla\nabla^{K3^Y}\lla\nabla^\ca{E}\lla\nabla^{\IZ_2}~.
\eqlabel{injections}$$
A point that is important is that this structure imposes a natural
coordinate system on the polyhedron. $\nabla^X$ is five-dimensional; within it is a
four-plane containing $\nabla^Y$; within this a three-plane containing $\nabla^{K3}$;
within this a two-plane containing $\nabla^\ca{E}$; and finally within this a line
corresponding to the zero dimensional \cy\ $\nabla^{\IZ_2}$.  

There is a further important property of this class of manifolds which is that there
is also a hierarchy of {\sl projections\/} that relate $X$ to $Z$: 
$$
\nabla^X\lra\nabla^Z\lra\nabla^{K3^Z}\lra\nabla^\ca{E}\lra\nabla^{\IZ_2}
\eqlabel{projections}$$
It is one of the observations of \cite{\Avrametal} that the roles of injections and
projections are interchanged by mirror symmetry so the mirror of each such
$X$ has also a structure analogous to \eqref{injections} and \eqref{projections} with
the replacements $Y\to \widetilde{Z}$ and $Z\to\widetilde{Y}$. Here and in the
following tildes are used to denote the mirror of a given manifold.

In these notes we shall be primarily concerned with the KLRY spaces and their
mirrors (these spaces will be discussed in details in \SS4).  
The polyhedra for these classes are very
different, the KLRY spaces have dual polyhedra that are small, typically with 20--70 points, while
their Newton polyhedra (which are the dual polyhedra of the mirrors) are large with typically
20,000--70,000 points.  The structure of these fourfolds suggests a precise
specification of the threefold $Z^X$, given $X$, by defining $Z$ in terms of its
mirror
$$ 
\F[X] = \Het[Z, V^X] \quad, \quad \widetilde{X} = (\widetilde{Z}, \cp1) \quad,
\eqlabel{mirrordual}$$
where in the second relation $\widetilde{X}$ and $\widetilde{Z}$ are the mirrors of
$X$ and $Z$ respectively. As we show in \SS2, the relation 
$\widetilde{X} = (\widetilde{Z}, \cp1)$, 
and hence $Z$, can be given a precise meaning
in virtue of the natural projections mentioned above.  For $X$ the KLRY space
$X_{mnp}$, this relation gives what we would expect $Z = Z_m = ({\cal E}, \IF_m)$,
the elliptic fibration of the Hirzebruch surface that is familiar from~ 
\Ref{\MorrisonVafa}{D. Morrison and C. Vafa, ``Compactifications
of $F$-theory on Calabi-Yau threefolds, I, II'',
\npb{473} (1996) 476, hep-th/9602114/9603161.}. 
However, for $X$ the mirror of
a KLRY space, the fact that we obtain a sensible definition of $Z$ in this way is far from trivial.

The occurrence of divisors that lead to superpotentials turns out to be a rather 
involved subject.  A first observation that is perhaps counterintuitive given
\cite{\Wittenone,\DGW} is that divisors of the fourfold with arithmetic
genus one are ubiquitous at least in the class of fourfolds that we study. (Of more
than 3000 manifolds all have these divisors and all their mirrors have them also.) 
What is less clear is how to deal with compactifications to four dimensions for which
we are interested in divisors of arithmetic genus one that are of the form
$\pi^{-1}(R)$ where $\pi$ denotes the projections $\pi: X\longrightarrow B^X$ onto
the base of the fibration and $R$ denotes a divisor of $B^X$.  Here the situation is
clearest when
$\pi^{-1}(R)$ consists of a single component which is a divisor of
$X$ of arithmetic genus one.  Frequently however, the preimage $\pi^{-1}(R)$ consists 
of several irreducible components with non-trivial intersection.  These preimages are
precisely the ones that give rise to the gauge group $G^X$ of the heterotic model.
\newpage
\section{X}{Construction of {\fourteenmath X}}
\vskip-30pt
\subsection{The KLRY Spaces}
The KLRY spaces, $X_{mnp}$, provide many examples of elliptically fibered fourfolds with
the structure \eqref{manifolds}. These spaces are defined by toric data 
 $$
Q_{mnp}~=~
\pmatrix{1&  1&  m&  0& p& 0& 2(m{+}p{+}2)& 3(m{+}p{+}2)& 0\cr
         0&  0&  1&  1& n& 0&   2(m{+}2)&   3(m{+}2)& 0\cr
         0&  0&  0&  0& 1& 1&        4&        6& 0\cr
         0&  0&  0&  0& 0& 0&        2&        3& 1\cr} \eqlabel{charges}$$
with the three integers having the ranges $1\le m,n \le 12$ and 
$0\le p\le \half n(m+2)$.
What is meant by this is, of course, that we have 9 coordinates 
$(\x_1, \x_2, \ldots,\x_9)$ that are identified under 4 scaling symmetries with
weights given by the rows of $Q$. Now the allowed monomials (those that have the same
multidegree as the fundamental monomial $\x_1\x_2\cdots \x_9$) are points in a
9-dimensional space. However in virtue of the scaling relations these points lie in a
5-dimensional plane. Thus the monomials associated with \eqref{charges} give rise to a
5-dimensional Newton polyhedron, $\D$. This polyhedron is reflexive and so corresponds
to a \cy\ fourfold.

Perhaps the simplest way to motivate the particular fourfold that we study is to
describe first a simple but inconsistent model associated with the manifold $X_{234}$
which we denote by $\widetilde X$ to save writing. The model is inconsistent for our
purposes since, despite the fact that it appears to be an elliptic fibration by
construction nevertheless, as we shall see, it fails to be an elliptic fibration. We
shall discuss this carefully in the following however the point at issue is whether the
embedding space $\IP^{\nabla}$ is an elliptic fibration. 
Such a fibration is expressed torically if every cone in the fan of $\widetilde{X}$
projects to some cone of the fan of the base. This is what fails for $\widetilde{X}$.
Repairing the fibration leads to a more complicated but viable model. 

To begin however consider then the dual polyhedron $\widetilde\nabla$ corresponding to
$\widetilde{X}$ shown below. The divisors $E_1$ and $\widetilde E$ are associated with 
an $SU(2)$ gauge group.  The fact that the groups are as seen from the polyhedron can be
verified by performing a calculation in the Cox coordinate ring.  In our case, this has
been checked by A.~Klemm.  As we
have already stressed, the fibration $\widetilde{X}\longrightarrow B$ is important for
us and   it is related to the projection onto the first three coordinates of the
points of $\widetilde\nabla$. 
\def\nablaXnought{%
\vbox{\offinterlineskip\tabskip=0pt\halign{%
\vrule height12pt depth 8pt\quad$##$\quad\hfil\vrule&
\hfil\quad(##&\hfil##,&\hfil##,&\hfil##,&\hfil##,&\hfil##&##)\quad\hfil\vrule&
\quad$##$\quad\hfil\vrule\cr
\noalign{\hrule}
\hfil\hbox{Vertex}&\multispan 7
\hfil$\mathstrut\hskip5pt\widetilde\nabla$\hfil\vrule&\hfil\hbox{Divisor}\cr
\noalign{\hrule\vskip3pt\hrule}
V_1&&-1&0&0&2&3&&\cr
V_2&&0&-1&0&2&3&&\cr
V_3&&0&0&-1&2&3&&\cr
V_4&&0&0&0&-1&0&&\cr
V_5&&0&0&0&0&-1&&\cr
{1\over 2}(V_3+E_1)&&0&0&0&2&3&&B\cr
{1\over 3}(V_2+V_4+V_7)&&0&0&1&1&2&&\widetilde E\cr
{1\over 2}(V_2+F)&&0&0&1&2&3&&E_1\cr
{1\over 2}(V_1+V_6)&&0&1&2&2&3&&F\cr
V_7&&0&1&3&2&3&&G\cr
V_6&&1&2&4&2&3&&Y\cr
\noalign{\hrule}
}}}
$$\nablaXnought$$
\subsection{Constructing the Fan for $\widetilde{X}$}
The polyhedron $\widetilde\nabla$ is not a simplex.  However, we can start by
taking the cones over its faces
$$\eqalign{
\{&V_2 V_4 V_5 V_6 V_7,~ V_2 V_3 V_4 V_5 V_6,~ V_1 V_2 V_3 V_4 V_5,~ 
V_1 V_2 V_3 V_4 V_6 V_7,\cr 
&V_1 V_2 V_3 V_5 V_6 V_7,~ 
V_1 V_3 V_4 V_5 V_6,~ V_1 V_2 V_4 V_5 V_7,~ V_1 V_4 V_5 V_6 V_7\}\cr} $$
Two of these cones $V_1V_2V_3V_4V_6V_7$ and $V_1V_2V_3V_5V_6V_7$ are not
simplicial.  They correspond to facets of $\widetilde\nabla$ that share a common 3-face
$V_1V_2V_3V_6V_7$.  We can see how to perform a triangulation by drawing the
$V_2V_3V_7F$-plane as well as the $V_2V_4V_7$ 2-face.  The following rules effect
the triangulation:
$$\eqalign{V_1V_6&\longrightarrow \{V_1F, FV_6\}\cr
V_2V_3V_7F&\longrightarrow \{V_2V_3B, V_2BE_1, V_2E_1V_7, BE_1V_7,BFV_7,V_3BF\}\cr
V_2V_4V_7&\longrightarrow 
\{V_2\widetilde EV_4, V_4\widetilde E V_7, V_7\widetilde EV_2\}\cr}$$ 

It is necessary to check that the $\widetilde{X}$ is actually an elliptic fibration, that
is, that the map $\pi : \widetilde{X}\longrightarrow B$ is smooth.  We may ensure this
by requiring that each cone of the fan for $\widetilde{X}$ should project onto some cone 
of the fan for the base $B$.  As we will show now, this is
\underbar{not} the case for $\widetilde\nabla$.
The problem arises in connection with the triangulation of the three face 
$(x_1, x_2, x_3, 2, 3)$ corresponding to $\Sigma^B$.  
In Figure\figref{fanfigs}, we draw the two plane $(0, x_2, x_3, 2, 3,)$ that lies within 
this face.  The cones of
$\Sigma^{\widetilde{X}}$ intersect this plane in the regions indicated.  The problem comes
from the cone from the origin of
$\widetilde\nabla$ (which is out of the plane) which interects this plane in the triangle
$V_2 E_1 G$.  If we project this cone onto the plane clearly it projects onto the
\underbar{union} of the cone generated by $V_1$ and $E_1$ and
the cone generated by $E_1$ and $G$.  The simplest way to fix the problem is to add the 
point
$E_2\sim (0, 0, 2, 2, 3)$ to
$\widetilde\nabla$ as in Figure \figref{fanfigs}b.  This divides the troublesome cone in 
two so that  the fan for $\widetilde{X}$ now projects nicely.
\vskip10pt
\def\fanfigs{\vbox{\vskip0pt\hbox{\hskip-10pt\epsfxsize=5.5truein\epsfbox{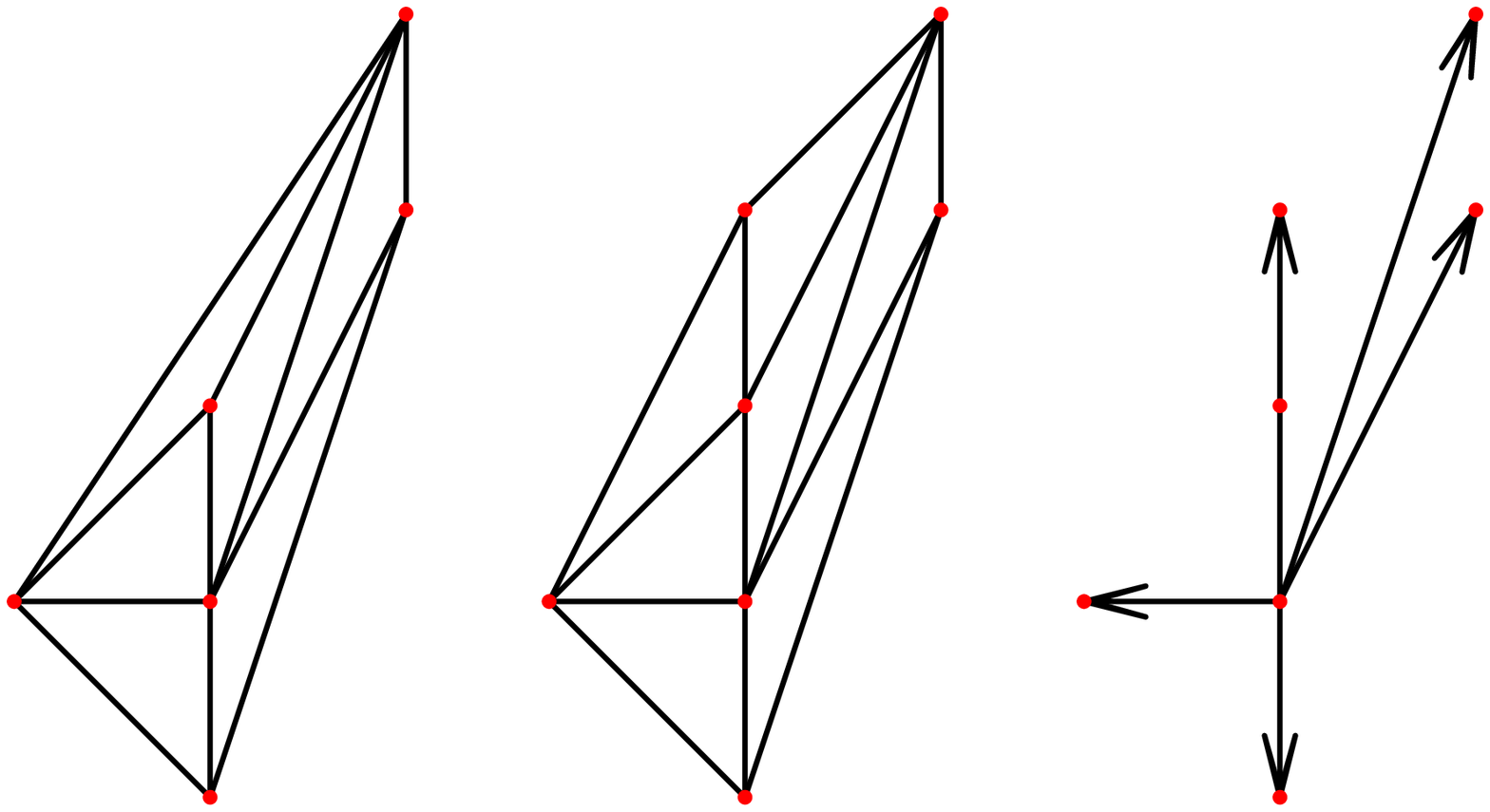}}}}
\figbox{\fanfigs}{\figlabel{fanfigs}}{The two-plane $(0,x_2,x_3,2,3)$ of $\nabla$. In (a) 
the plane is given for $\widetilde{X}$. The bad cone is subdivided in (b) corresponding to
the improved manifold $X$. The cones now project properly onto the fan~(c).}
\place{0.25}{2.35}{$Y^-$}
\place{1.4}{1.6}{$C_1$}
\place{2.05}{3.7}{$F$}
\place{2.05}{4.4}{$G$}
\place{4.05}{2.35}{$Y^-$}
\place{5.2}{1.6}{$C_1$}
\place{5.85}{3.7}{$F$}
\place{5.85}{4.4}{$G$}
\place{4.75}{3.0}{$E_1$}
\place{4.75}{3.7}{$E_2$}
\vskip10pt
When we take the convex hull of $\widetilde\nabla\cup \{E_2\}$ we find that the 
new polyhedron contains also the point $E_3\sim (0, 0, 1, 1, 1)$.  Moreover, the
point $\widetilde E$ now lies in the interior of a codimension one face.  In this
way, the $SU(2)$ associated with $\{E_1, \widetilde E\}$ is replaced by a $G_2$
associated with $\{E_1, E_2, E_3\}$. 

At the level of practical calculation, note that the preimage of the ray
$(0,0,1)$ of the base of $\widetilde{X}$ is the divisor $E_1 + \widetilde E$.  It follows that
$(E_1 + \widetilde E)^4$ should vanish since the intersection calculation pulls
back from the intersection calculation on the base.   This consistency check
fails for the fan for $\widetilde{X}$. 

We are not yet quite done with the changes. In order to enforce this condition, we
have to add the point $C_2\sim (1, 0, -1, 1, 2)$. The data for our consistent
fourfold is displyed in Table~4.1.  Note that $\{C_1, C_2\}$
correspond to an additional $SU(2)$ gauge group. We have added the divisor $C_2$,
which could have been omitted, in order to show how we may build up the group. We make
some further comments about how to build up the groups in \SS4.3 below.  Table 4.1
summarizes the polyhedron and divisors for $X$.
\midinsert
$$
\vbox{\def\skip{\hskip5pt}\def\space{\hphantom{1}}\def\nostar{\hphantom{^*}}
\offinterlineskip\tabskip=0pt\halign{%
\vrule height11pt depth7pt\skip$#$\skip\hfil\vrule&\hfil\skip$#$\skip\vrule&
\hfil\skip(#&\hfil#,&\hfil#,&\hfil#,&\hfil#,&\hfil#&#)\skip\hfil\vrule&
\skip$#$\skip\hfil\vrule\cr
\noalign{\hrule}
\omit\vrule height18pt depth14pt\titlebox{Relation}{to vertices}\hfil\vrule
&\chi\hfil&\multispan 7
\hfil$\mathstrut\hskip5pt\nabla^X$\hfil\vrule&\hfil\hbox{Divisor}\cr
\noalign{\hrule\vskip3pt\hrule}
V_1&0\nostar&&-1&0&0&2&3&&Y^+\cr
V_2&0\nostar&&0&-1&0&2&3&&Y^- = F+G\cr
V_3&1^*&&0&0&-1&2&3&&C_1\cr
V_9&1^*&&0&0&-1&1&2&&C_2\cr
V_4&-89\nostar&&0&0&0&-1&0&&2H + E_3 - C_2\cr  
V_5&-368\nostar&&0&0&0&0&-1&&3H + E_3 - C_2\cr 
{1\over 2}(V_3 {+} E_1)&1^*&&0&0&0&2&3&&B\cr
{1\over 2}(V_3 {+} 2E_2)&1^*&&0&0&1&2&3&&E_1\cr
{1\over 2}(V_1{+}V_6)&1^*&&0&0&2&2&3&&E_2\cr
{1\over 2}(V_5{+}E_2)&0\nostar&&0&0&1&1&1&&E_3{=}C_1+C_2-(E_1+2E_2+2F+3G+4Y^{+})\cr
V_7&1^*&&0&1&2&2&3&&F\cr
V_8&1^*&&0&1&3&2&3&&G\cr
V_6&0\nostar&&1&0&4&2&3&&Y^{+}\cr
\noalign{\hrule\vskip3pt\hrule}
\multispan{10}\vrule height15pt depth10pt\hfil 
~$h^{11}=8~,~h^{31}=2897~,~h^{21}=1~,~h^{22}=11662~,~\ch_E=17472~
$\hfil\vrule\cr
\multispan{10}\vrule height15pt depth10pt\hfil
$H=B+C_1+C_2+E_1+E_2+2F+2G+2Y^{+}$\hfil\vrule\cr
\noalign{\hrule} 
}}$$
\centerline{{\bf Table 4.1:} The divisors for the manifold $X$.}
\vskip10pt
\endinsert
\midinsert
\vbox{
\def\Bfig{\vbox{\vskip0pt\hbox{\hskip0pt\epsfxsize=4.5truein\epsfbox{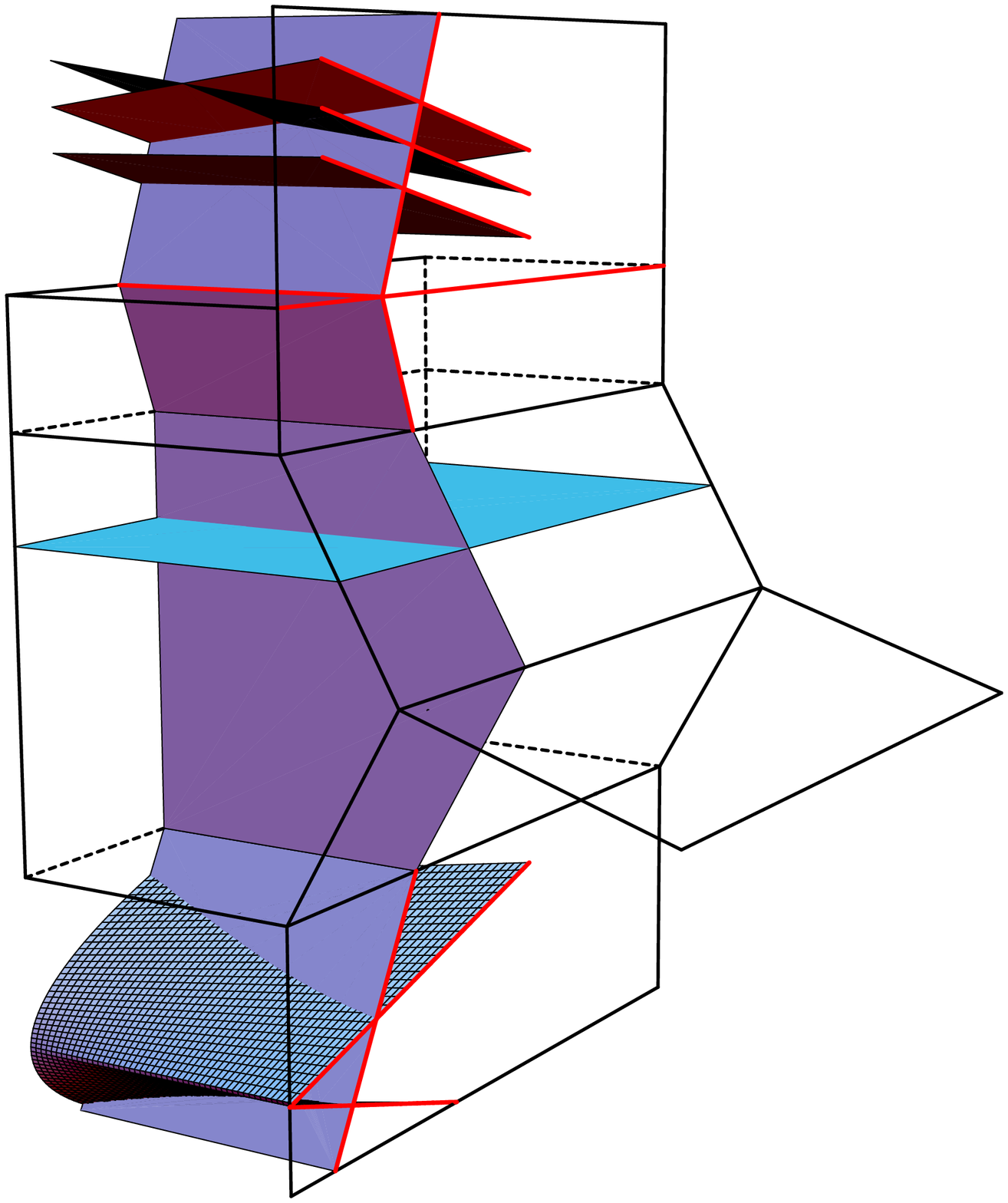}}}}
\figbox{\Bfig\vskip-30pt}{\figlabel{Bfig}}{A sketch of how the various divisors
intersect the base $B$ of the elliptic fibration showing also the degenerate fibers
corresponding to the groups $SU_2$ and $G_2$. The surfaces shown as $E_3Y^{+}$ are really a single
connected surface. The `components' that are shown meet in pairs. The surface $C_2Y^{+}$ is ruled by
quadrics which degenerate, exceptionally, into a pair of lines. This explains the $1/2$ that appears in the
relation $\ell^3 = \half C_2FY^{+}$.}
\place{1.05}{5.03}{$B^Z$}
\place{1.6}{4.53}{$Y^+$}
\place{5}{6.23}{$\matrix{\nwarrow\hskip-10pt\cr &G\cr &\downarrow\cr}$}
\place{4.2}{5.28}{$\matrix{&\hskip-8pt B\cropen{-3pt} \swarrow\cr}$}
\place{3.8}{3.33}{$\matrix{&&\nearrow\cr\leftarrow&F\cr}$}
\place{4.0}{6.13}{$E_1E_2G$}
\place{1.9}{6.2}{$E_2Y^+$}
\place{0.9}{6.8}{$E_3Y^+\hskip3pt\left\{\vrule height15pt depth5pt width0pt\right.$}
\place{1.8}{2.48}{$C_1Y^+$}
\place{0.9}{3.23}{$C_2Y^+$}
}
\endinsert
\noindent
>From \cite{\KLRY} we also see that 
 $$\displaylines{
h^{k,0}(E_1)=h^{k,0}(E_2)= h^{k,0}(C_j)= h^{k,0}(F)=h^{0,0}(G)=0~\hbox{for}~k=1,2,3\cr
\hbox{and}~h^{0,0}(E_i)= h^{0,0}(C_j)= h^{0,0}(F)=h^{0,0}(G)=1~, 
\hbox{while}~~h^{2,0}(E_3)=0~~\hbox{and}~~h^{1,0}(E_3)=1~.\cr}$$
For our model $X$ our construction of the fan proceeds similarly to the case of
$\widetilde{X}$.  We begin with the cones over the facets of the polyhedron that is the
convex hull of the vertices $V_1, V_2, \ldots, V_8$.  This yields the cones
$$\eqalign{
\{&V_4 V_5 V_6 V_7 V_8,~ V_2 V_3 V_4 V_5 V_6,~ V_1 V_3 V_4 V_5 V_6,~ 
V_1 V_2 V_3 V_4 V_5,~ V_1 V_2 V_3 V_4 V_6 V_7 V_8,\cr
 &V_1 V_2 V_3 V_5 V_6 V_7 V_8,~ V_1 V_4 V_5 V_6 V_7,~
  V_1 V_2 V_4 V_5 V_8,~ V_1 V_4 V_5 V_7 V_8,~ V_2 V_4 V_5 V_6 V_8\}\cr}  $$
There are two facets which are not simplices, $V_1V_2V_3V_4V_6V_7V_8$ and
$V_1V_2V_3V_5V_6V_7V_8$, and these facets have a common 3-face
$V_1V_2V_3V_6V_7V_8 = V_1V_6\,Y^{-}C_1FG$ which corresponds to the fan for the base $B$
of the  elliptic fibration (see Table~4.1).  To see how to perform a triangulation 
we make reference to
\fig\figref{fanfigs} which depicts the two-plane $V_1V_2V_3V_6V_7V_8$ that lies within 
this three-face.  By associating divisors to the points of the figure and noting
that $V_1$ and $V_6$ are the only points of the three plane that do not lie in the
two plane and that these points are joined by a line that passes through $F$. We see
that the following rules effect the triangulation:
 $$
\eqalign{V_1V_6&\longrightarrow \{V_1F, FV_6\}\cr
V_2V_3V_7V_8F&\longrightarrow \{V_2V_3B, V_2BE_1, V_2E_1V_8, E_1V_7V_8,
BE_1V_7,BFV_7,V_3BF\}\cr 
V_5V_8&\longrightarrow \{V_5E_3, E_3V_8\}\cr}~.$$

This yields a simplicial fan.  Finally, we insert $V_9$ which lies in the cone
$V_3V_4V_5$ by means of the rule:
$$V_3V_4V_5\longrightarrow \{V_3V_4V_9, V_4V_5V_9, V_3V_5V_9\}~.$$
The result is a fan of 54 cones $\Sigma^X = V_1\Sigma^Y\cup V_6\Sigma^Y$, where
$\Sigma^Y$ denotes the fan of 27 cones 
$$\eqalign{ 
\{&B E_1 V_2 V_4,~ B F C_1 V_4,~ B V_2 C_1 V_4,~ E_3 E_1 V_2 V_5,~ 
B E_1 V_2 V_5,~ B F C_1 V_5,~ B V_2 C_1 V_5,~ E_3 V_2 V_4 V_5,\cr 
&B F V_4 G,~ B E_1 V_4 G,~ B F V_5 G,~ E_3 E_1 V_5 G,~ B E_1 V_5 G,~ E_3 V_4 V_5 G,~ 
F V_4 V_5 G,~ E_3 E_1 V_2 E_2,~\cr 
&E_3 V_2 V_4 E_2,~ E_1 V_2 V_4 E_2,~ E_3 E_1 G E_2,~ E_3 V_4 G E_2,~ E_1 V_4 G E_2,~ 
F C_1 V_4 C_2,~ V_2 C_1 V_4 C_2,~\cr 
&F C_1 V_5 C_2,~ V_2 C_1 V_5 C_2,~ F V_4 V_5 C_2,~ V_2 V_4 V_5 C_2\}\cr}  $$  
and $V_1\Sigma^Y$
denotes the set obtained by appending $V_1$ to each cone of $\Sigma^Y$ (and
similarly for $V_6\Sigma^Y$).
\pageinsert
\vbox{
\def\KY{\vbox{\vskip0pt\hbox{\hskip50pt\epsfysize=4truein%
\epsfbox{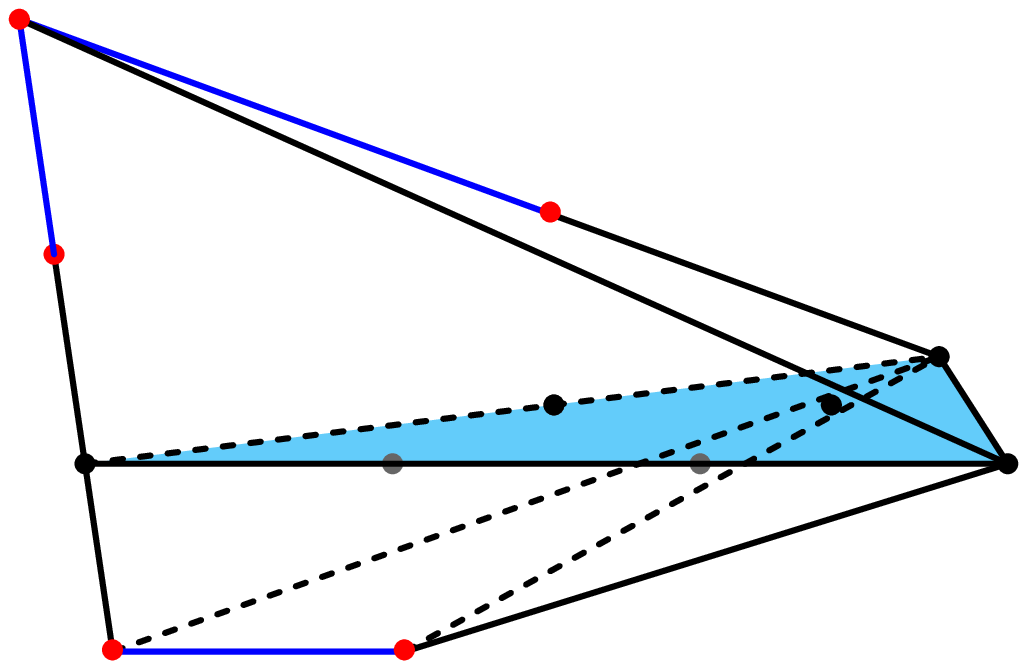}}}}
\figbox{\KY\vskip-60pt}{\figlabel{KY}}{The polyhedron for $K^Y$. The horizontal
triangle $\nabla^\ca{E}$ divides the polyhedron into a top and a bottom.
The extended Dynkin diagrams corresponding to the groups $G_2$ and $SU(2)$ are
seen as the blue lines.}}
\vfil
\vbox{
\def\triangs{\vbox{\vskip10pt\hbox{\hskip50pt\epsfysize=2truein%
\epsfbox{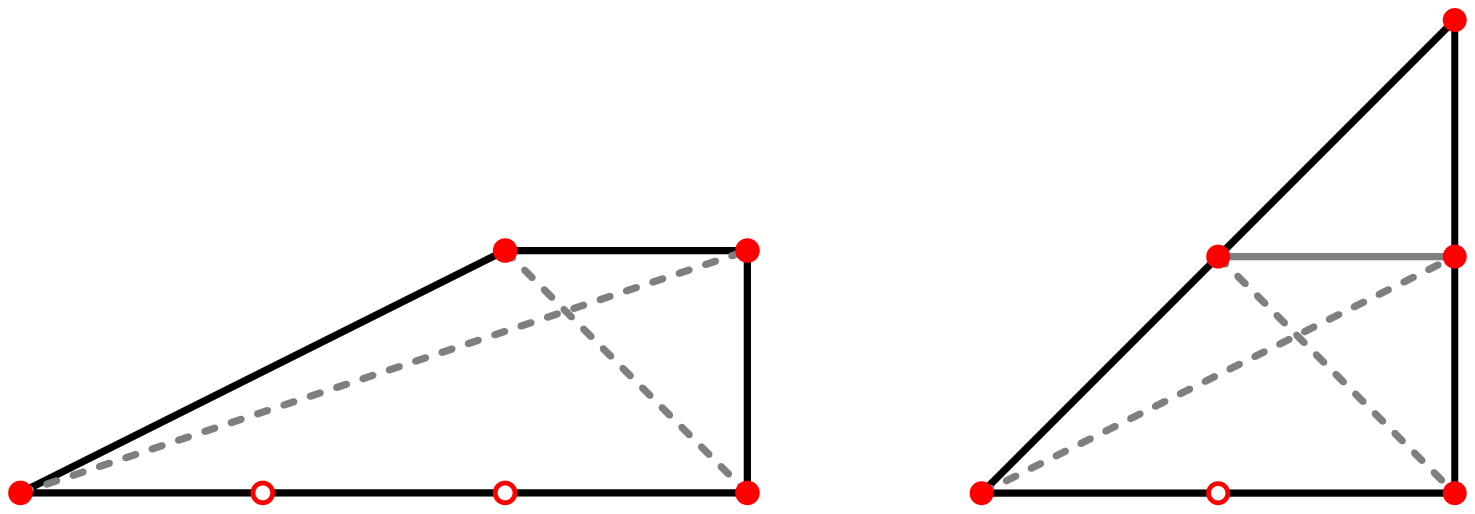}}}}
\figbox{\triangs\vskip0pt}{\figlabel{triangs}}{The two faces of $\nabla^{K^Y}$ that 
admit more than one triangulation.}}
\vfil
\endinsert
We can in fact express the combinatorics of the fan in a better way. To do this
we write each cone as a product and a fan as a sum of cones. Thus with this 
understanding
$\Sigma^Y = B E_1 V_2 V_4 + B F C_1 V_4 + \cdots + V_2 V_4 V_5 C_2$ and 
$\S^X = (V_1+V_6)\S^Y$. Now with this notation we may express $\S^Y$ in the form
 $$
\S^Y = (F+V_2)\S^C + (G+V_2)\S^E + FG\S^\ca{E}$$
where $\S^C$ and $\S^E$ denote 
the fans over the faces of the half-polyhedra corresponding to the $SU(2)$ and $G_2$.
and $\S^\ca{E}$ denotes the fan over the faces of $\nabla^\ca{E}$.
 $$\eqalign{
\S^C&= B C_1 V_4 + C_1 C_2 V_4 + B C_1 V_5 + C_1 C_2 V_5 + C_2 V_4 V_5\cr
\S^E&= E_1 E_2 E_3 + B E_1 V_4 + E_1 E_2 V_4 + E_2 E_3 V_4 + B E_1 V_5 + 
E_1 E_3 V_5 + E_3 V_4 V_5\cr
\S^\ca{E}&= B V_4 + B V_5 + V_4 V_5~.\cr} $$
Could we have chosen different fans? The answer is yes however if we restrict to 
fans that project to the fan for $B^X$ then there are just four choices and these
are related by flops corresponding to the fact that two of the faces of
$\nabla^{K^Y}$ each admit two different triangulations as in
Figure~\figref{triangs}. The yukawa couplings for these  four fans are however the
same which indicates that the flops affect the embedding space but not the
hypersurface~$X$.
\subsection{Extending the Groups}
It is of course also possible to extend the gauge group by adding points 
to $\nabla$. We think of this as building up the half-polyhedra that 
project down onto the divisors of $B$ and wish to show how this may be
implemented on the fan. Suppose we begin by seeking to build up a half-polyhedron
over the divisor $F$. We change notation by denoting $F$ by $F_1$, so that the
new points are $F_2$, $F_3$ etc. and by denoting $G$ by $G_1$ since we will
want also to add a group over $G$. Our starting point is the fan
 $$
\S^Y = (F_1+V_2)\S^{\hbox{bot}} + (G_1+V_2)\S^{\hbox{top}} + F_1G_1\S^\ca{E}$$ 
One checks that the new points $F_2$, $F_3$,... all lie in the cone
$F_1V_4V_5$ of $\S^Y$. This cone occurs as a face in two of the terms in
$\S^Y$. It occurs both in $F_1\S^{\ca{E}}$ and $F_1\S^C$ since both $\S^C$
and $\S^{\ca{E}}$ contain the cone $V_4V_5$. Let $\S^{F_1}$ denote the trivial
half-polyhedron over $F_1$, that is the half-polyhedron with no extra points,
and denote by $\S^{F}$ the half-polyhedron corresponding to the new group. We
will write $\S^{F_1}\longrightarrow \S^F$ for the process of extending the fan
to the fan over the faces of the new half-polyhedron. 
 
For the term $F_1\S^{\ca{E}}$ we observe that
 $$
F_1\S^{\ca{E}} = \S^{F_1} \longrightarrow \S^F~.$$
While for the term $F_1V_4V_5$ in $F_1\S^C$ we note that
 $$
F_1V_4V_5 = \S^{F_1} - F_1B(V_4 + V_5)$$
so we have
 $$\eqalign{
F_1\S^C &= F_1\left(\S^C - C_2V_4V_5\right) + C_2(F_1V_4V_5)\cr
        &= F_1\left(\S^C - C_2V_4V_5\right) + C_2\left( \S^{F_1} -
F_1B(V_4+V_5)\right)\cr
        &\longrightarrow F_1\left(\S^C - C_2V_4V_5\right) + C_2\left( \S^F -
F_1B(V_4+V_5)\right)~.\cr}$$
In this way we see that $\S^Y\longrightarrow \widetilde{\S}^Y$ with
 $$
\widetilde{\S}^Y = (F_1 + V_2)\S^C + (G_1 +V_2)\S^E +(G_1 + C_2)\S^F
-F_1C_2\S^{\ca{E}}~.$$
Note that the function of the last term is to remove terms that are present in
$F_1\S^C$ so there are really no minus signs in this expression. 

Now let us add a group over $G_1$. The new points all lie in the cone
$G_1V_4V_5$ which is contained in the terms $G_1\S^F$ and $G_1\S^E$. In $\S^F$
there is a single cone of the form $F_jV_4V_5$ and we denote this divisor $F_j$
by $F_{\hbox{max}}$. For $G_1\S^F$ we write
 $$\eqalign{
G_1\S^F &= G_1( \S^F - F_{\hbox{max}} V_4V_5 ) + 
F_{\hbox{max}}\left( \S^{G_1} - G_1B(V_4 + V_5)\right)\cr
        &\longrightarrow G_1( \S^F - F_{\hbox{max}} V_4V_5 ) +
F_{\hbox{max}}\left( \S^G - G_1B(V_4 + V_5)\right)\cr
        &= G_1\S^F + F_{\hbox{max}}\S^G - F_{\hbox{max}}G_1\S^{\ca{E}}~.\cr}$$
Similarly
 $$
G_1\S^E \longrightarrow G_1\S^E + E_3\S^G - E_3G_1\S^{\ca{E}}~.$$
In this way we arrive at a fan 
 $$\eqalign{
\widetilde{\widetilde{\S}} &= (F_1 + V_2)\S^C + (G_1 + V_2)\S^E + (G_1 + C_2)\S^F
+ (E_3 + F_{\hbox{max}})\S^G\cr 
&\qquad - (E_3G_1 + F_{\hbox{max}}G_1 + C_2F_1)\S^{\ca{E}}~.\cr}$$
\subsection{New groups}
The groups may be extended by adding points to the polyhedron and we have carried out this procedure to
extend the group $G_2$. A non-simply laced group results from the effect of monodromy on a simply laced
group. Some of the divisors for the simply laced group are identified under the monodromy. The resulting
divisor(s) of the non-simply laced group are the ones that may have $\ch\neq 1$. It seems that while it is
frequently the case that these divisors that result from identification under monodromy have $\ch\neq 1$
that this is not always the case. Our first extension $G_2\subset SO(7)$ leads to a group all of whose
divisors have $\ch=1$. however as we extend the group further to $SO(9)$, $F_4$, $SO(11)$ and $SO(13)$ we
find that these cases all exhibit dissident divisors. The data is summarized by the following table.
$$
\def\skip{\hskip7pt}
\def\titlebox#1#2{\lower7pt\hbox{\hsize=.75in\vbox{\vskip5pt\centerline{#1}\vskip5pt\centerline{#2}}}}
\vbox{\offinterlineskip\tabskip=0pt\halign{%
\vrule height12pt depth 8pt%
\hfil\skip$#$\skip\hfil\vrule&\hfil\skip$#$\skip\hfil\vrule
&\hfil\skip$#$\skip\hfil\vrule&\hfil\skip$#$\skip\hfil\vrule&\hfil\skip$#$\skip\hfil\vrule
&\hfil\skip$#$\skip\hfil\vrule&\hfil\skip$#$\skip\hfil\vrule&\hfil\skip$#$\skip\hfil\vrule\cr
\noalign{\hrule}
\omit\vrule height25pt depth18pt\hfil\skip\titlebox{Apparent}{Group}\hfil\skip\vrule
& h^{11} & h^{31} & h^{22} & h^{21} & \d & ~\bar{\d}~  
& \titlebox{$\ch$~of\vphantom{p}}{dissident(s)}\cr
\noalign{\hrule\vskip3pt\hrule}
\omit\vrule height2pt\hfil\vrule &&&&&&&\cr
G_2    &  8 & 2897 & 11662 & 1 & 0 & 0& \{0\}\cr
\noalign{\hrule}
SO(7)  &  9 & 2895 & 11660 & 0 & 0 & 0& \hbox{no dissidents}\cr
\noalign{\hrule}
SO(9)  & 10 & 2894 & 11660 & 0 & 2 & 0& \{3\}\cr
F_4    & 10 & 2894 & 11660 & 0 & 4 & 0& \{3,3\}\cr 
\noalign{\hrule}
SO(10) & 11 & 2869 & 11564 & 0 &  0 & 0 & \hbox{no dissidents}\cr
SO(11) & 11 & 2869 & 11564 & 0 & 12 & 0 & \{13\}\cr
\noalign{\hrule\vskip3pt\hrule}
SO(13) & 13 & 2787 & 11244 & 0 & 25 & 0& \{26\}\cr
\noalign{\hrule}
E_{7b} & 14 & 2790 & 11236 & 12&  0 & 0& \{-11\}\cr
\noalign{\hrule}
E_8    & 15 & 2825 & 11292 & 56&  0 & 0& \{-55\}\cr
\noalign{\hrule}
}}$$
We will discuss in \SS 6 the contribution to the superpotential
from the corresponding divisors; here we make some further comments on the table:
\subsubsection{$G_2$:~}The dissident divisor, $E_3$ corresponds to the end of the Dynkin diagram
\lower2pt\hbox{\epsfxsize=0.7truein\epsfbox{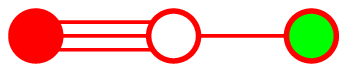}} away from the extending root. This divisor
is the one that corresponds to the three nodes of the Dynkin diagram for $SO(8)$ that are identified under a
$\IZ_3$ monodromy. The dissident divisor contributes one to $h^{21}$, and so also to $h^{12}$ and hence
contains two three-cycles.
\smallskip
\subsubsection{$SO(7)$:~}All the divisors have arithmetic genus unity even though the group is not
simply laced.
\smallskip
\subsubsection{$SO(9)$:~}There is a dissident divisor
\hbox{\epsfxsize=1truein\epsfbox{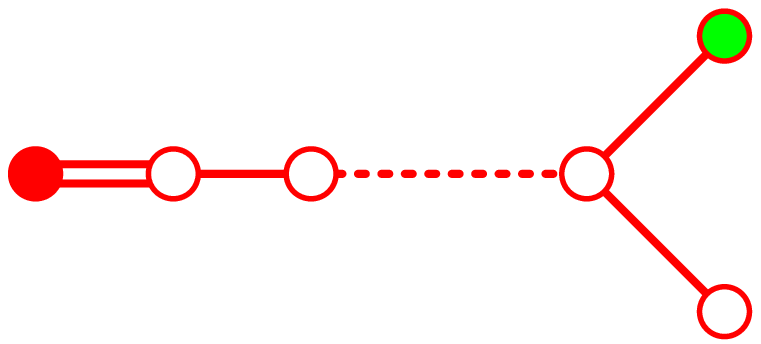}} again corresponding to the end of the Dynkin
diagram  away from the extending root. This is the node that corresponds to the two nodes of the
$SO(10)$ that are identified under a $\IZ_2$ monodromy. In this case the dissident divisor
contributes to $\d$, the number of non-toric parameters.
\smallskip
\subsubsection{$F_4$:~}There are now two dissidents at the end of the Dynkin diagram
\lower1.3pt\hbox{\epsfxsize=1truein\epsfbox{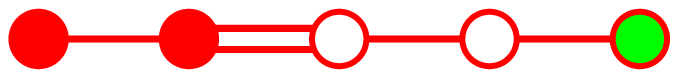}} away from the extending root which correspond
to the nodes of the $E_6$ that are identified under a $\IZ_2$ monodromy. The
dissidents contribute to $\d$. The Hodge numbers of the manifolds corresponding to $SO(9)$ and $F_4$ are the
same suggesting that they are in fact the same manifold. The number of non-toric parameters is less for
$SO(9)$ than for $F_4$ suggesting that $SO(9)$ gives the better description and that the true group is
perhaps $SO(8)$.
\smallskip
\subsubsection{$SO(10)$:~}We include this group for purposes of comparison with $SO(11)$. The group
$SO(10)$ is simply laced so all the nodes of the Dynkin diagram correspond to divisors with
arithmetic genus unity. For this case there are no non-toric parameters.
\smallskip
\subsubsection{$SO(11)$:~}This case is similar to $SO(9)$. There is a dissident divisor which
contributes to $\d$. The Hodge numbers for this manifold are the same as those for $SO(10)$ which
suggests that under a generic deformation the group becomes $SO(10)$.
\smallskip
\subsubsection{$SO(13)$:~}This case is interesting and exhibits a new phenomenon which it has in
common with the following two examples. The group is not simply laced nevertheless all the nodes of
the Dynkin diagram correspond to divisors with arithmetic genus unity. The dissident divisor, in
the toric description, arises not from the group per se but because the half polyhedron that projects
down to a ray in the fan of $B^X$ contains the divisor $\{0,0,1,0,0\}$ as a point interior to a facet.
This divisor is associated with a further blow-up of the base. 

More precisely: after resolving the general singularities of the Weirstrass model, which gives rise
to the $S(13)$ configuration, the fourfold is still singular; the curve of singularities is
a rational curve. It turns out that after one blow up the fourfold $X$ is smooth and still satisfies
the Calabi-Yau condition; $X$ is elliptically fibered over a threefold $B$, which is the blow up of
$B^X$ (the common base of the other examples) along a rational curve $\Gamma$. The fiber of this new
fibration are all curves, and there are no new gauge groups.
  
The elliptic threefold over the surface $ B^Y$, obtained after resolving the general singularities
over $\sigma _\infty$ is singular at one point, after blowing up this point to another surface $B$
and normalizing we obtain a new Calabi-Yau, mapping to $B$, with one dimensional fibers.
\smallskip
\subsubsection{$E_8$:~}Again the nodes of the Dynkin diagram correspond to divisors with arithmetic
genus unity and the dissident divisor is as in the previous case.

A more careful statement is again that after resolving the general singularities
of the Weirstrass model, which gives rise to the $E_8$ configuration,
the fourfold is still singular and further blow ups are needed.
It turns out that after all the necessary blow ups the fourfold
$X$  still satisfies the Calabi-Yau condition; $X$ is elliptically
fibered over a threefold $B$, which is the blow up
of $B^X$ (the common base of the other examples) along a curve $\Gamma$.
The fiber of this new fibration are all curves, and
there are no new gauge groups. We can also consider the elliptic threefold over the surface
$B^Y$ ($B^X$ is fibered by $B^Y$, the Calabi-Yau fourfold is fibered by such Calabi-Yau threefolds).
$B^Y$ is the Hirzebruch surface $\IF_3$ blown up at a point. We denote by $f$ the
exceptional divisor of this blow up, $g$ the strict transform of a fiber of the fibration
$\IF_3\to \IP^1$ and $\sigma _\infty$ the section with negative self-intersection. Our models
have a
$SU(2)$ gauge group over a section $\sigma _0$, such that $\sigma _0 \cdot g=0, \ \sigma_0 \cdot
f=1$. A quick check shows that the general elliptic fibration with an $E_8$ gauge group over 
$\sigma_\infty$, acquires extra singularities over $9$ different points $P_i, \ i=1, \cdots 9$,  at
the intersection of $\sigma _\infty$ with the divisor of $I_1$ singular fibers. These $9$ points are
the intersection of $\Gamma$ and $B^Y$. It can be easily verified that the resolution of these
singularities introduces $9$ new divisors on the threefold: if we want to mantain
the equidimensionality conditions (all the fibers being curves), then we have to blow up the base at
each $P_i$. The new threefold is Calabi-Yau and there are no new gauge groups. 
\smallskip
\subsubsection{$E_{7b}$:~}This is one of the ways of realizing $E_7$ and is similar to the two cases
above. All the nodes of the Dynkin diagram correspond to divisors with arithmetic genus unity. The
dissident divisor is $\{0,0,1,0,0\}$ which is a point interior to a facet.  
\newpage
\section{yuk}{The Yukawa Couplings and Mori Cone}
\vskip-30pt
\subsection{The Yukawa Couplings}
The topological yukawa couplings $D_iD_jD_kD_l$ are calculated from the fan by means
of the program SCHUBERT. The most important vanishing relations in the intersection
ring can be read off directly from the fan or more simply from \fig\figref{fanfigs}.
We see, for example, that $V_1$ and $V_6$ always lie in different cones and that $G$
never occurs in the same cone with $C_1$ or $C_2$ and that $F$ never occurs in the
same cone with any of the $E_i$. It follows that these divisors do not intersect in
$\IP_\nabla$ and hence do not intersect in $X$. In this way we learn that
 $$\displaylines{
GC_i=0~,~~~FE_j=0,~~~C_iE_j=0~,~~~BC_2=BE_2=BE_3=0\cr
F(F+G) = G(F+G) = 0~,~~~Y^2 = 0~,~~~(3H+E_3)E_2 = 0\cr}$$
where the last of these identities follows from the fact that $V_5$ and $E_2$ never
lie in the same cone. Four further quadratic identities follow by examining the intersection numbers
 $$
E_1E_3 = 0~,~~~BH = 0~,~~~E_1H = 0~,~~\hbox{and}~~(2H - C_2)C_1 = 0~.$$
Taken together with the previous identities these furnish a basis for the quadratic
relations between the divisors.
The nonzero intersection numbers are given below for a slightly redundant basis that includes also
the divisor $E_3$: 
$$
\def\row#1#2#3#4#5{#1&#2&#3&#4&#5\cr}
\vbox{\offinterlineskip\tabskip=0pt\halign{%
\vrule height15pt width0pt$#$\quad\hfil
&\quad$#$\quad\hfil&\quad$#$\quad\hfil&\quad$#$\quad\hfil&\quad$#$\hfil\cr
\row{B^4= -82}{B^3 C_1= 48}{B^3 E_1= 4}{B^3 F= 8}{B^3 Y= 7}
\row{B^2 C_1^2= -28}{B^2 C_1F= -6}{B^2 C_1 Y= -4}{B^2 E_1^2= -10}{ B^2 E_1 G= 2}
\row{B^2 E_1 Y= 1}{B^2 F^2= 2}{ B^2 F G= -2}{B^2 F Y= -1}{ B^2 G^2= 2}
\row{B^2 G Y= -1}{ B C_1^3= 16}{B C_1^2 F= 4}{ B C_1^2 Y= 2}{B C_1 F Y= 1}
\row{B E_1^3= 24}{B E_1^2 G= -4}{B E_1^2 Y= -3}{B E_1 G Y= 1}{B F^3= -4}
\row{B F^2 G= 4}{B F^2 Y= -1}{B F G^2= -4}{B F G Y= 1}{B G^3= 4}
\row{B G^2 Y= -1}{C_1^4=-144}{ C_1^3 C_2= 192}{C_1^3 F=8}{C_1^3 Y= 8}
\row{C_1^2 C_2^2= -272}{C_1^2C_2 F= -16}{ C_1^2 C_2 Y= -12}{C_1^2 F Y= -2}{C_1 C_2^3= 384}
\row{C_1 C_2^2 F= 24}{C_1 C_2^2 Y= 16}{C_1 C_2 F Y= 2}{C_2^4= -528}{ C_2^3 F= -32}
\row{C_2^3 Y= -20}{C_2^2 F Y= -2}{E_1^4=-56}{E_1^3 G= 8}{ E_1^3 Y= 8}
\row{E_1^2 E_2^2= 2}{ E_1^2 E_2 Y= -1}{E_1^2 G Y= -2}{ E_1 E_2^3= 4}{E_1 E_2^2G= -2}
\row{E_1 E_2^2 Y= -1}{E_1 E_2 G Y= 1}{E_2^4= -48}{E_2^3 E_3= 36}{E_2^3 G= 8}
\row{E_2^3 Y= 8}{E_2^2 E_3^2= -18}{ E_2^2 E_3 G= -6}{E_2^2E_3 Y= -9}{ E_2^2 G Y= -2}
\row{E_2 E_3^2 Y= 9}{E_2 E_3 G Y= 3}{E_3^4= -72}{E_3^3 G= 24}{E_3^2 G Y= -6~.}
}} $$
\subsection{The Mori Cone of $X$}
The Mori cone (what is this and why do we even care) of the embedding space
$\IP_{\nabla}$ may be found by the method of positive piecewise linear functions
as explained in KLRY (see also~
\Ref{\OdaPark}{T. Oda and H. S. Park, 
``Linear Gale Transforms and Gel'fand-Kapranov-Zelevinskij
Decompositions'', Tohoku Math. J. (2) 43 (1991), no. 3, 375--399.}).  
Recall that a dimension is added to the vector space defined by the points of $\nabla$
and a 1 is prepended to each of the points corresponding to the divisors (so that
$D_0\sim (1, 0, 0, 0, 0)$, etc.). The cones of the fan extend to simplicial cones whose
vertex is at the origin of this extended space. 

A piecewise linear function $m$ is defined on the extended space so as to be
linear on each cone.  If $u$ is a point in a cone $\sigma$ with generators $u_i$,
$i\in I$, then $u$ can be expressed uniquely in the form
$$\vec{u} = \sum_{i\in I}\lambda_i \vec{u}_i\quad;\qquad\lambda_i\ge 0~.$$
The function $m(\vec{u})$ is given in terms of $m_i = m(\vec{u}_i)$ by
$$ m(\vec{u}) = \sum_{i\in I} \lambda_i m_i$$
which is equivalent to giving a vector $\vec{m}_{\sigma}$ on each cone such that
$$m(\vec{u}) = \vec{m}_{\sigma}\cdot \vec{u}\quad{\rm for}\quad
\vec{u}\in\sigma~.$$

The function $m$ is a positive piecewise linear function if
$$\eqalign{m(\vec{u}) &= \vec{m}_{\sigma}\cdot\vec{u}\quad;\quad u\in\sigma\cr
       m(\vec{u}) &\ge\vec{m}_{\sigma}\cdot\vec{u}\quad;\quad
u\notin\sigma\cr}\eqlabel{ineqs}$$
The system of inequalities \eqref{ineqs}, being linear, is specified by the
coefficients that appear.  In this way, they specify a cone which is identified
with the Mori cone of $\IP_{\nabla}$.  The integral basis for the system
\eqref{ineqs} may be identified with the generators of the Mori cone. (This process is 
performed in detail for the simple case of the threefold $Z$ in appendix A.) For fourfolds
this process requires computer calculation.

For the case at hand,  the relations between the divisor classes may be used to
express the system \eqref{ineqs} entirely is terms of a basis consisting of the
divisor classes 
 $$
\{B,~ C_1,~ C_2,~ E_1,~ E_2,~ F,~ G,~ Y\}~.\eqlabel{basis}$$  
In other words, the $m_i$ corresponding to the remaining divisors may be set to
zero.  The system \eqref{ineqs} then consists of 83 inequalities which are
generated by the following coefficient vectors
\def\acone{\eqalign{
&a^{-1}\hskip-6pt= (\- 1,-2,\- 2,\- 0,\- 0,  -1,\- 0,\- 0 )~,\cropen{5pt}
&a^0 = (\- 1,  -1,\- 1,  -1,\- 0,\- 0,\- 0,\- 0 )~,\cropen{5pt}
&a^1 = (  -1,\- 1,\- 0,\- 0,\- 0,  -1,\- 1,\- 0 )~,\cropen{5pt}
&a^2 = (\- 0,\- 0,\- 0,  -1,  -1,\- 0,\- 1,\- 0 )~,\cropen{5pt}
&a^3 = (  -1,\- 0,\- 0,\- 1,\- 0,\- 1,  -1,\- 0 )~,\cropen{5pt}   
&a^4 = (\- 0,\- 0,\- 0,\- 1,  -2,\- 0,\- 0,\- 0 )~,\cropen{5pt}
&a^5 = (\- 0,\- 0,\- 0,\- 0,\- 1,\- 0,\- 0,\- 0 )~,\cropen{5pt}
&a^6 = (\- 0,\- 1,  -1,\- 0,\- 0,\- 0,\- 0,\- 0 )~,\cropen{5pt}
&a^7 = (\- 1,  -3,\- 3,\- 0,\- 0,\- 0,\- 0,\- 0 )~,\cropen{5pt}
&a^8 = (\- 1,\- 0,\- 0,  -2,\- 0,\- 0,\- 0,\- 0 )~,\cropen{5pt} 
&a^9 = (\- 0,\- 0,\- 0,\- 0,  -2,\- 0,\- 0,\- 1 )~,\cr}}
\vskip10pt
 $$\acone$$
\vskip10pt
\noindent
The vectors on the right are the coefficients of the inequalities that generate
\eqref{ineqs}. We may associate them also with the curves that generate the Mori 
cone. The components of the vectors are also the intersection numbers
 $$
a^i{}_j~=~a^i\cdot D_j$$
of the curves with the divisors $D_j$ of the basis \eqref{basis}. The Mori
cone that we have obtained has nine edges $a^i,~i=1,\ldots,9$. There are in addition
two further generators $a^{-1}$ and $a^0$ which are internal to the cone and which
are required because in this case the edges do not generate the cone
$$\eqalign{
&a^0\hskip-3pt= \half (a^6 + a^7 + a^8)~,\cropen{5pt} 
&a^{-1} ~= \half (a^0 + a^4 + a^7)~.\cr}$$
Note that since there are nine edges rather than eight the cone is not simplicial.
A ninth edge is however necessary since the relation of linear dependence is 
 $$
a^1 - a^2 - 3a^3 + 4a^4 + 2a^5 ~=~ 0 $$
and so no edge can be written as a positive combination of the others. 

The complication is that what we have calculated is the Mori cone of the 
embedding space $\IP_\nabla$ rather than the Mori cone of $X$. The procedure
advocated by Cox and Katz\cite{\CoxKatz} is to 

\item{(i)~}compute {\sl all\/} the possible fans for
$\IP_\nabla$ that is triangulate $\nabla$ in all possible ways. The program
PUNTOS can accomplish this in cases that are not too complicated. 
\smallskip
\item{(ii)~}Compute using the program SCHUBERT the yukawa couplings
$Y^{ijkl} = D_iD_jD_kD_l$ corresponding to each fan. Fans that lead to
different couplings $Y^{ijkl}$ correspond to different phases of the theory in
the sense of the linear sigma-model. Fans that lead to the same $Y^{ijkl}$
correspond to different resolutions of the embedding space that do not affect
the \cy\ hypersurface $X$. That is the corresponding $\IP_\nabla's$ are
related by flopping curves that do not intersect $X$. Thus the fans should be
grouped into classes classified by the $Y^{ijkl}$.
\smallskip
\item{(iii)~}Within a given class the Mori cones for the embedding spaces will
in general be different however the true Mori cone, \ie the Mori cone for
$X$, should be contained in each of them. Thus we proceed, for a given
class, by computing the intersection of all the corresponding Mori cones of
the embedding spaces.

We have carried through this program for our~$\nabla$; though we shall see that the resulting cone is
still too large. A total of 990 fans were found which when classified by the yukawa couplings fall
into 7 classes. These 7 classes turn out to correspond to the 7 ways of triangulating the
$(0,x_2,x_3,2,3)$ plane shown in \fig\figref{fanfigs}. 
The class corresponding to our couplings, \ie the Table at the beginning of this
section, comprises 20 fans all of which correspond to the triangulation of the
$(0,x_2,x_3,2,3)$ plane shown in the central figure in
\fig\figref{fanfigs}. Although they coincide in this plane these 20 fans
are different leading to 20 distinct Mori cones. Of the 20 only 4 satisfy the
condition that was discussed previously that each cone of the fan of $\IP_\nabla$
project onto some cone of the fan for $B$. The other fans in this class correspond
to the same yukawa couplings as the fans that do satisfy the projection criterion
so it must be the case that although these fans do not realize $\IP_\nabla$ as an elliptic
fibration they do realize $X$ as an elliptic fibration. 

The task of finding the intersection of the twenty cones can be done by means
of a computer program however in this particular case it is easy to do by
hand. It happens that among the twenty cones there are many edges that appear
in a certain cone and then with opposite sign in another cone. In such a case
it is easy to see that the intersection of the two cones is contained in the
cone formed by discarding the edges that appear with opposite sign and taking
the union of the remaining edges for the two cones. Thus we may discard all
the edges that appear with opposite sign and take the union of all the
remaining edges. In this way we see that the intersection must be contained
in the following cone that has the ten edges $\ell^i,~i=1,\ldots,10$
\smallskip
$$\def\skip{\hphantom{\half}}
\eqalign{
&\ell^0=(\- 1,  -1,\- 1,  -1,\- 1,\- 0,\- 0,\- 0)= a^0 + a^5 \cropen{3pt}
&\ell^1=(\- 0,\- 0,\- 0,  -1,  -1,\- 0,\- 1,\- 0)=\skip E_1E_2Y^{+}\hskip-3pt
=a^2\cropen{3pt}
&\ell^2=(\- 1,  -2,\- 2,\- 0,\- 0,\- 0,\- 0,\- 0)=\skip C_1FY^{+}=a^6 + a^7\cropen{3pt}
&\ell^3=(\- 0,\- 1,  -1,\- 0,\- 0,\- 0,\- 0,\- 0)=\half C_2FY^{+}=a^6\cropen{3pt} 
&\ell^4=(\- 1,\- 0,\- 0,  -2,\- 1,\- 0,\- 0,\- 0)=\skip E_1GY^{+}=a^5 + a^8\cropen{3pt}
&\ell^5=(\- 0,\- 0,\- 0,\- 1,  -2,\- 0,\- 0,\- 0)=\skip E_2GY^{+}=a^4\cropen{3pt} 
&\ell^6=(\- 0,\- 0,\- 0,\- 0,\- 1,\- 0,\- 0,\- 0)={1\over 3}E_3GY^{+} =a^5\cropen{3pt}
&\ell^7=(  -1,\- 1,\- 0,\- 0,\- 0,  -1,\- 1,\- 0)=\skip FBY^{+}\hskip5pt=a^1\cropen{3pt}   
&\ell^8=(  -1,\- 0,\- 0,\- 1,\- 0,\- 1,  -1,\- 0)=\skip GBY^{+}\hskip4pt=a^3\cropen{3pt} 
&\ell^9=(\- 0,\- 0,\- 0,\- 0,  -2,\- 0,\- 0,\- 1) =\skip E_1E_2G=
{1\over 3}E_2E_3G\hskip4pt=a^9\cropen{3pt} 
&\ell^{10}\hskip-5pt=(  -1,\- 0,\- 1,\- 0,\- 0,\- 0,\- 0,\- 0)= 
a^0 + a^1 + a^3\cropen{0pt} 
}\eqlabel{mcone} $$
It is easy to see also that each of the edges of this cone is contained in
each of the twenty cones with which we started so that it is in fact the
intersection we were seeking. Again in this case the edges do not generate 
the cone and we require also the internal generator
 $$
\ell^0~=~\half(\ell^2 + \ell^4 + \ell^6) $$
This cone contains the true Mori cone and by
inspection we identify curves of $X$ with each of the generators apart from
$\ell^0$ and $\ell^{10}$. Thus the edges $\ell^1,...,\ell^9$ are true edges.
Notice also that the divisors of our basis appear in the curves as
 $$\eqalign{
\{C_1,~C_2\}FY^{+}=\{C_1,~C_2\}Y^{+}Y^{-}~&,~~~~~\{E_1,~E_2,~E_3\}GY^{+}=\{E_1,~E_2,~E_3\}Y^{+}Y^{-}~,
\cropen{5pt}
\hbox{and}~~&\{F,~G\}BY~.\cr}$$
It is interesting to note that if we compute the intersection matrix between
the divisors $C_i$ and the curves $C_jFY^{+}$ with $i,j=1,2$ we find
 $$
C_iC_jFY^{+} ~=~ \pmatrix{ -2&\- 2\cr 
                         \- 2&  -2\cr}$$
which we recognise as the extended Cartan matrix for $SU(2)$ by which we mean
the matrix corresponding to including the extending root of the algebra. This is in
accord with the observations of Intriligator {\it et al.\/}
\Ref{\Intriligatoretal}{K. Intriligator, D. R. Morrison and N. Seiberg,
``Five-Dimensional Supersymmetric Gauge Theories and Degenerations of \cy\ Spaces'',\\
\npb{497} (1997) 56, hep-th/9702198.}. 
If we consider the intersections between the
divisors $E_i$ and the curves $\ell^j$ with $i = 1,2,3$  and $j=4,5,6$ then we find the
extended Cartan matrix corresponding to $G_2$
 $$
E_i\ell^j ~=~ \pmatrix{  -2&\- 1&\- 0\cr
                       \- 1&  -2&\- 3\cr
                       \- 0&\- 1&  -2\cr}~.$$
 
The curve $\ell^9 = E_1E_2G$ maps to the \cp{1}\ of the base of
\cy-fibration $X=(Y,\cp1)$ while $\ell^1$ maps to a fiber of $B^Z$. 
Since there are ten edges there are two linear relations between them one
of these involves $\ell^{10}$ the other relation is more interesting and
permits the elliptic fiber to be expressed in two different ways
 $$
\ca{E} = FGY = 
(1, 0, 0, 0, 0, 0, 0, 0) = \ell^2 + 2\ell^3 = \ell^4 + 2\ell^5 + 3\ell^6~.
\eqlabel{fiber}$$

We can show that $\ell^0$ and $\ell^{10}$ cannot be generators of the true
cone. Consider first $\ell^{10}$. If this is a generator of the Mori Cone
then it is an irreducible curve that is contained in $B$, since
$\ell^{10}\cdot B = -1$, and which also intersects $C_2$, since 
$\ell^{10}\cdot C_2 = 1$. But this is impossible since $B$ and $C_2$ do not
intersect. In a similar way we see that $\ell^0$ is not a true generator
since such a curve would have to be contained in both $C_1$ and $E_1$ which
however do not intersect.

Note now that the generators
$\ell^1,\ell^2,\ell^4,\ell^5,\ell^7,\ell^8,\ell^9$, that is the generators
$\ell^1,\ldots,\ell^9$ with $\ell^3$ and $\ell^6$ omitted, define a
seven-plane $L$, say, within the eight dimensional cone. For any vector $k$
in the lattice we can define a height relative to $L$
 $$
h(k) = \det(k,\ell^1,\ell^2,\ell^4,\ell^5,\ell^7,\ell^8,\ell^9) = h.k $$
where on the right $h$ denotes the vector $(6,8,5,4,2,8,6,4)$. Now
$h(\ell^0) = 1$, $h(\ell^3)=3$, $h(\ell^6)=2$ and $h(\ell^{10})=-1$ so
$\ell^{10}$ lies on one side of $L$ and $\ell^0$, $\ell^3$ and $\ell^6$ lie
on the other side. This seems to provide a counterexample to the conjecture of Cox and Katz (see also~
\cite{\Szendroi}) that the procedure we have followed should yield the true Mori cone. 

We have seen that
$\ell^{10}$ is not a true generator and the question arises as to whether we can discard all the
points that have negative height with respect to $L$. Now we have to express $\ell^0$ as
a positive integral combination of generators, since $\ell^0$ cannot be a
generator, and since $h(\ell^0) = 1$ which is less than $h(\ell^3)$ and
$h(\ell^6)$ we see that we must have at least one generator with negative
height. 
\subsection{Volumes of the Divisors}
There are two natural ways to parametrize the \K-form. The first is to write
it directly in terms of the basis of divisors
 $$
J~=~tB + s_1C_1 + s_2C_2 + s_3E_1 + s_4E_2 + s_5F + s_6G + vY~.$$
in this expression $t=J\cdot\ca{E}$ is the volume of the elliptic fiber.
Another useful parametrization of the
\K-form is obtained by taking the volumes of the curves $\ell^1,...,\ell^9$ (defined in
\eqref{mcone}) as coordinates. We set
 $$
J\cdot\ell^i~=~
\left(\m_{+},~\g_1,~\g_2,~\e_1,~\e_2,~\e_3,~\d_F,~\d_G,~\m_{-} \right)~,~~~i=1,...,9~.
\eqlabel{params}$$
There are 9 parameters on the right so there is a linear relation which is a
consequence of \eqref{fiber}
 $$
t~=~\g_1+2\g_2~=~\e_1+2\e_2+3\e_3~.\eqlabel{volE}$$

We wish now to examine the relation between the volumes of the b-divisors and
the volumes of the Mori generators. This is of interest since the
superpotential arises, in the M-theory description, through contributions of
the form $\exp(2\p i \vol(D))$ while in the heterotic description it arises
through instanton corrections and the instantons are linear combinations of
the Mori generators.
Now 
 $$
\vol(D)~=~{1\over 3!} J^3 D $$
which is cubic in the parameters of $J$ while the volumes of the Mori curves
and so of course the instantons are linear in the parameters of $J$. The
volumes of the b-divisors are for the most part complicated cubic expressions in the
parameters. The simplest of these expressions are those for the volumes of
$E_1$ and $E_2$
 $$\eqalign{
\vol(E_1) &=  \e_1 \left( \m_{-}\m_{+} + \e_1(\half\m_{-}+\m_{+}) + {4\over 3}\e_1^2
\right)\cropen{7pt} 
\vol(E_2) &=  \e_2\m_{-}\left( \m_{+} + \half\e_2 \right)~.\cr}$$
It seems that one should consider the limit with $t$ small. Since the
parameters that we have introduced through \eqref{params} are all positive
it follows in virtue of \eqref{volE} that the $\g_i$ and $\e_j$ tend to zero
with $t$. We expect $\vol(D)$ to tend to zero linearly with $t$ for every
b-divisor $D$ so the neglect of terms of $\ca{O}(t^2)$ leads to expressions
with a term of $\ca{O}(t)$ as a factor. For $F$ we have in this limit
$\vol(F)\asymp t\m_{-}\d_F$. The corresponding linearized expressions for the
other b-divisors are as follows

$$\def\skip{\hphantom{2}}
\eqalign{
\vol(C_1) &\asymp \skip\g_1 (2 \d_F + 3 \d_G + \m_{+}) (4 \d_F + 4 \d_G + \m_{-})\cropen{3pt}
\vol(C_2) &\asymp 2\g_2 (2 \d_F + 3 \d_G + \m_{+}) (4 \d_F + 4 \d_G + \m_{-})\cropen{3pt}
\vol(E_1) &\asymp \skip\e_1\m_{-}\m_{+}\cropen{3pt}
\vol(E_2) &\asymp \skip\e_2\m_{-}\m_{+}\cropen{3pt}
\vol(E_3) &\asymp 3\e_3\m_{-}\m_{+} \cropen{3pt}
\vol(F)~  &\asymp \skip t\d_F(\m_{-} + 2\d_F + 4\d_G)\cropen{3pt}
\vol(G)~  &\asymp \skip t\d_G(\m_{-} + 2\d_G)~. \cr} $$
\newpage
\section{superpot}{The Superpotential}
\vskip-30pt
\subsection{Characterization of the Divisors Contributing to the Worldsheet Instantons}
Let us denote with $\pi : X \to B^X$ the elliptic fibration, $p: B^X
\to B^Z$ the fibration by rational fibers (generally $\IP^1$),
and by $ \e : X \to B^Z$ the composed K3-fibration.
We assume that $p$ is equidimensional (replacing, if necessary, $B^Z$ by
a suitable blow up). The divisors that contribute to the superpotential
are the divisors $D$ such that $\p(D)$ is a divisor, $C$, of $B^X$ and 
$p (\pi (D))$ is a curve, $\g$, in $B^Z$:
\vskip10pt
$$\matrix{
X\cropen{3pt} 
\hskip3pt\downarrow\raise3pt\hbox{\hskip-4pt$\scriptstyle{\p}$}
&\searrow\raise3pt\hbox{\hskip-7pt$\scriptstyle{\e}$}\cropen{3pt}
B^X&\hskip-5pt\hbox{$\rightarrow\hskip-9pt {\vrule height10pt
width0pt}^p$}&B^Z\cr}
\hskip1in %
\matrix{
D\cropen{3pt} 
\hskip3pt\downarrow\raise3pt\hbox{\hskip-4pt$\scriptstyle{\p}$}
&\searrow\raise3pt\hbox{\hskip-7pt$\scriptstyle{\e}$}\cropen{3pt}
C&\hskip-5pt\hbox{$\rightarrow\hskip-9pt {\vrule height10pt
width0pt}^p$}&\g\cr}
$$
\vskip10pt
Most the divisors that we construct via the toric construction 
contribute to space-time instantons.
It turns out that divisors contributing to
the worldsheet instantons are nicely divided in 3 different
types, each of which has a distinct meaning in physics:

\noindent {\bf (a)} If $D$ {\bf does not correspond
to a non-abelian gauge group} and $D= \epsilon ^*(\gamma)$, where
$\g = \epsilon (D)$. 
In this case $1=\chi(D) =- \gamma ^2$.
In fact, $$\chi (D)=D^2 \cdot c_2 /24= -\gamma ^2 \cdot \chi _{top} (S)
/24= - \gamma ^2,$$ where $S$ is the general fiber of
the $K_3 $-fibration $X \to B^Z$. 
Furthermore, the adjunction formula shows that 
$\gamma$ is a smooth rational curve. Such curve is necessarily an edge of
the Mori cone
of $B^Z$.

\noindent {\bf (b)} If $D$ {\bf does not correspond
to a non-abelian gauge group} and $D \neq \epsilon ^*(\gamma)$,
then $p : B^X \to B^Z$ is not a $\IP^1$-bundle; not all such divisors
contribute to the superpotential, even if $\chi ({\cal O} _D)=1$ (see also~
\Ref{\DiaconescuRajesh}{D. Diaconescu and G. Rajesh, 
``Geometrical Aspects of Five-Branes
in Heterotic / F Theory Duality in Four-Dimensions'', 
JHEP {\bf 9906} (1999) 002, hep-th/9903104}). 
HOwever the divisors $F$ and $G$  do contribute(see Section 4).

\noindent {\bf (c)} If $D$ {\bf corresponds to a gauge group},
 $D$ arises from a degeneration of the K3 fiber $S$: therefore $D$ generates a
{\it non-perturbative} gauge group and the corresponding heterotic model will
have singularities.
 In particular $\epsilon (D)= \gamma$ is a component of the discriminant locus
(with gauge groups) of the elliptic fibration $Z \to B^Z$.
 We have the following 2 cases:
 
\noindent {\bf $\bullet$} c.1) $C= \pi (D) = p^*(\g)$

\noindent {\bf $\bullet$} c.2) $C$ is a component of $ \pi (D) \subset p^*(\g)$
\subsection{Comparison with the Heterotic Superpotential}
 At this point we do not know much about cases b) and c), so we are
concentrating on the divisors of type a). Most of the explicit examples
in the literature are of type a).
 The corresponding heterotic superpotential (via the F-theory/heterotic
duality)
is expected to
be a function of the volume of the curves $p(\pi (D))$ and thus to be linear.
 While we do not see any mathematical a priori reason of why this should
be true, it turns out to be so in all the examples examined hitherto. 
Typically the divisors contributing to the superpotential on $F$-theory
compactifications are 
finite in number: this is because $B^X$, the base of the elliptic fibration
needs to have an effective first Chern class ($c_1(B^X) \geq 0$);
in the toric case also the number is always finite\ 
\Ref{\Grassi}{A. Grassi, 
``Divisors 0n Elliptic \cy\  Four Folds and the Superpotential in F Theory.
1'', J. Geom. Phys. {\bf 28} (1998) 289}.
 
The examples in~
\Ref{\Mayr}{P. Mayr,``Mirror Symmetry, $N=1$ superpotentials
and Tensionless Strings on Calabi-Yau Four-Folds'', hepth/9610162.}\ 
and \cite{\Grassi} have only a finite number of
divisors and these are of type (a) and a simple computation shows that
the superpotential is linear in the volumes of these divisors, up to an
overall factor.
The computation is more complicated in the case where there are
infinitely many divisors, as in the following examples:
\subsection{Andreas' Examples}
In his paper\ 
\Ref{\Andreas}{B. Andreas, ``N=1 Heterotic / F Theory Duality'',\\
Fortsch. Phys. {\bf 47} (1999) 587, hep-th/9808159.}\  
on heterotic/F-theory duality
Andreas considers certain examples closely related to the one in \cite{\DGW}:
there are infinitely many divisors contributing to the superpotential.
We summarize his argument here:
\subsubsection{The threefold $B_n = B^X$.}
$S \to \IP^1$ is $\IP^2$ blown up at 9 points and $ \IF_r \to \IP^1$ a
Hirzebruch surface. $B_n = B^X$ is the fiber product of $S$ and $\IF_r$
with base $\IP^1$, with $2r=n$.
 In particular $B^X$ is a $\IP^1$-bundle over $S$, but is not a product
unless,  $r=0$ (where $B^X = S \times \IP^1$, as in \cite{\DGW}).
\subsubsection{The threefold $\calb _r$.}
Let $X^3 _n \to \IF_n$, $n=2r$ be a smooth Calabi-Yau 3-fold, with an
involution $\tau$ compatible with the involution on $\IP^1$: $ z \to {-z}$.
(Such threefolds can be obtained by choosing appropriate coefficients for
the Weierstrass model.)
Set $X^3_n / \t = \calb_r $; $\calb_r$ is a smooth threefold. There is a
natural elliptic fibration $ \calb_r \to \IF_r$.
(Note: if $r=n=0, \calb_0 = B$, as in \cite{\DGW}.)
\subsubsection{The fourfold $X^4_n=X_r$.}
$X_r$ is the fiber product of $S \to \IP^1$ and $ \calb _r  \to \IP^1$.
By construction $X_r$ is elliptically fibered over $B^X$, while is fibered
by K3 surfaces over $S= B^Z$ (the basis of the heterotic dual.

 It can be verified that  $X_r$ is a Calabi-Yau 4-fold, with heterotic dual
$X^3_n$.
(If $r=0$, then $X _0$ is the Weierstrass model of the $X$ in \cite{\DGW};
the calculation in \cite{\DGW} computes also the superpotential for $X_0$,
up to a
factor.) We are interested in the contribution to the superpotential from
worldsheet
 instantons.
\subsubsection{The divisors contributing to the superpotential}
The divisors contributing to the superpotential via worldsheet instanton
are, as in
\cite{\DGW} the inverse images of the section of the fibration $S \to \IP^1$.
As in \cite{\DGW}, they are all isomorphic to $\calb_r$.
If $r>0$, there are other divisors contributing to the superpotential
via spacetime instantons (some correspond to gauge groups). 

Denote by  $\{\Gamma_0,  \Gamma_1 , \cdots, \Gamma_s \}$  the generators
of $H^2(B^X,\IR)$ and by 
 $$
\{ B, \pi^*(\Gamma_0 ), \cdots, \pi^*(\Gamma_s ), \L_1,\ldots,\L_w \}$$
the  generators $H^2(X,\IR)$,  where $B \sim B^X$ and the $\L_j$ 
correspond to gauge groups (there are the exceptional
divisors of the morphism to the Weierstrass model). Without loss of generality we take
$\L_j\cdot B=0$ and take $\Gamma _0 \sim S \sim B^Z$ and $\Gamma _i = p^*(\gamma _j)$,
where $\{ \gamma_j \}$ generate the \K\  cone of $S$.
As before we can express the \K\ form as a linear combination of the divisors
 $$
J = tB + \widetilde{\G}(u) + \L(v)~,~~~\hbox{with}~~
\widetilde{\G}(u) = \sum_j u^j\pi^*(\Gamma_j)~~\hbox{and}~~\L(v) = \sum_k
v^k\L_k~.$$
The volume form for a threefold is then 
 $$
{1\over 3!}J^3 = {1\over 3!}\left( tB + \widetilde{\G}(u) + \L(v)\right)^3$$
We will  now compute the volume of a {\it worldsheet instanton\/},
that is,  $\vol(D_\gamma)$ for $D_\gamma = \pi^*(p^*\gamma)$,
with $\gamma$ a section of the elliptic fibration $B^Z \to \IP^1$. 
We use the following facts: $B^2= B_{|B}= K_B $, and $-K_{B^Z}$ is  a fiber
of the elliptic fibration $B^Z \to \IP^1$. By the adjunction formula
$B^2 =K_B= -2 S - (r+1) p^*(F)$, where $F$ is a fiber of the fibration $S
\to \IP^1$
 and $B^3 = {K_B \cdot K_B }_{|B}$.
Note that again we have $t= \vol{\cal E}$, and that, if $f$ is the homology
class of the 
$\IP^1$ bundle $p: B^X \to S$, $ t_0 = 
(\sum t_j \pi^*\Gamma_j) \cdot f = S \cdot f = \vol f.$
Using the geometry of the fiber products involved we see that:
 $$\eqalign{ 
B^2\cdot\widetilde{\G}(u)\cdot D_\g~=&~ -(r-1)\vol_{B}(f) - 2\,\vol_S(\gamma)\cropen{5pt}
B^3 D~=&~ 4 \cropen{5pt}
\widetilde{\G}(u)^3 \cdot D_\gamma~=&~ 0\cropen{5pt}
\L(v)^3\cdot D_\gamma ~=&~ d (r,v)  \cropen{5pt}
B^2 \cdot \L(v)\cdot D_\gamma~=&~  0\cropen{5pt}
\L(v) \cdot \widetilde{\G}(u)^2 \cdot D_\gamma ~=&~ 0\cropen{5pt}
B \cdot \L(v)^2 \cdot D_\gamma~=&~ 0  \cropen{5pt}
\widetilde{\G}(u)\cdot  \L(v)^2 \cdot D_\gamma~=&~ 
c(r,u,v)\,\vol_{B^X}(f) + c'(r,u,v) \,\vol_S (\gamma)  \cropen{5pt}
\widetilde{\G}(u)^2 \cdot B \cdot D_\gamma~=&~ 
2\,\vol_S(\gamma) \, \vol_{B^X}(f) -r\,\vol _{B^X}^2(f) \cropen{5pt}
B \cdot  \L(v) \cdot \widetilde{\G}(u)~=&~ 0 ~,\cr} 
$$

\noindent where $d(r,v),  c(r,u,v), c'(r,u,v)$ are linear function on $r$, which do not
depend on the choice of $\gamma$,  and are zero for $r=0$.
Then:
$$ \exp(-\vol(D))= \exp(A) \times \exp\left[-C\,\vol(\g)\right],$$
where $A$ and $C$ are the same for every divisor $D_\g$. 
This function, up to a constant depending only on
$r$, is the expression in \cite{\DGW}
(and for $r=0$ is equal to this expression); 
then, up to a constant the superpotential is as in~\cite{\DGW}. 
\subsection{Comparison with $d=3$ Dimensional Yang-Mills Theory}
Katz and Vafa in \cite{\KatzVafa,\Vafatwo} consider divisors contributing
to the superpotential arising from resolution of singularities
of the Weierstrass model; the divisors are associated to a simple
gauge group $G$. They assume that there is no adjoint matter.
By the chain of duality in \cite{\Vafatwo},
$F$-theory compactified on a circle is dual to $M$-theory compactified on
the Calabi-Yau manifold $X$; the radius of the circle is the inverse to
the K\"ahler class of the elliptic fiber.
 In this way, one obtains a $N=2$ theory with $d=3$; Katz and Vafa
show that, under this duality the $F$-theory superpotentials
become the expected superpotential. 
 A key point in their computation is that each
divisor $D_i$ corresponding to the nodes of the affine Dynkin diagram
of the Group $G$ contribute to the superpotential, that
is they satisfy the conditions  (2.4) and (2.5).
 For examples, in the simply laced cases, one can write 
$[\ca{E}]=\sum_{i=1}^{r+1}a_i [e_i]$, where $\ca{E}$ is the class of the
elliptic fiber,
$e_i$ is the fiber of each $D_i$ (a ruled surface) and $a_i$ is the Dynkin
index of the corresponding node of the Dynkin diagram).
\subsection{New Features for a Non-Simply Laced Group}
 While the argument implied by the chain of dualities should imply the same
conclusion, we are unable to make the argument work for the
cases where not all the divisors satisfy the condition that $\ch(\ca{O}_D)=1$, as in the example in
Sections 5 and 6. These mixed configurations, in which some but not all of the nodes of the Dynkin
diagram contribute to the superpotential, correspond to a genuine instability since in these cases it
is not possible to satisfy the conditions $dW=0$ corresponding to the supersymmetric
vacuum states. 

If   $\chi ({\cal O}_D) >1$, then
$h^2 (D) >0$.
It follows  that $D$ is not general in the moduli of
$X$, that is the locus for which $D$ deforms in the family $\cal X$
(of complex deformation of $X$) is a complex submanifold of codimension 
$h^2(D)$.  The argument needed is a modification of the corresponding
statement for Calabi-Yau threefolds  in~ 
\Ref{\Wilson}{P. M. H. Wilson, ``The Kaehler Cone on Calabi-Yau Threefolds'',\\
Inventiones mathematicae {\bf 107} (1992).}.
In this case  $h^2(D)$ contributes to the number of non-toric parameters.
We will argue in this case that the Calabi-Yau fourfold is
non-general and the usual techniques for counting the divisors
contributing to the superpotential do not suffice. For example,
there is evidence that one should also consider the
contribution of reducible divisors.

If  $\chi  ({\cal O}_D) \leq 0$, then $h^1 >0$; in the
our examples $h^2(D)=0$ and  there are no non-toric parameter,
so the divisor $D$ will be effective (with the same Hodge number),
for all points of the complex moduli space of $X$.
Following \cite{\KLRY} we see that in the toric case
 $$
h^{2,1}(X)=\sum _{ q \in dim {\tilde \theta} =2} (1- \chi _q)= \sum h^1(D)~.$$
This gives an interesting, yet-little studied structure on the heterotic dual~
\Ref{\Diaconescu}{D. Diaconescu, Private comunication.}.
  
By construction the Calabi-Yau fourfold $X$ is fibered
by the family of Calabi-Yau threefolds $Y$. 
 It follows that the locus for which $D_Y$ deforms in the family $\cal Y$
(of complex deformation of $Y$) is a complex submanifold of codimension 
$h^1(D_Y)$. At this point we are not sure of the implication of this fact.
\subsection{The prefactor.}
Ganor~
\Ref{\Ganor}{O. J. Ganor, ``A Note on Zeros of Superpotentials in F Theory'',\\
\npb{499} (1997) 55, hep-th/9612077.}\  
argued that the contribution of a divisor $D$ to the superpotential
is multiplied by a pre-factor $f$. In most cases,
the pre-factor is non-zero; Ganor gives a necessary and sufficient
condition for this prefactor to vanish. Interestingly
this can happen only when $h^{2,1}(X)>0$. So one would hope
that in the Dynkin diagram configurations with ``dissident divisors" the
prefactor
would actually be zero for the non dissident ones.
An easy computation in the case of $G_2$, \SS 4.4, shows that the
prefactor is nevertheless non-zero for the divisors contributing to
the superpotential.
\newpage
\chapno=-1
\section{appZ}{Geometry of {\fourteenmath Z}}
\vskip-30pt
\subsection{The Divisors}
The polyhedron that we obtain by deleting the third column of $\nabla^X$ has the 
structure
\vskip10pt
\def\divisorsZ{%
\vbox{\offinterlineskip\tabskip=0pt\halign{%
\vrule height12pt depth 8pt\quad$##$\quad\hfil\vrule&\hfil\quad$##$\quad\vrule&
\hfil\quad(##&\hfil##,&\hfil##,&\hfil##,&\hfil##&##)\quad\hfil\vrule&
\quad$##$\quad\hfil\vrule\cr
\noalign{\hrule}
\omit\vrule height18pt depth14pt\hfil\titlebox{Relation}{to vertices}\hfil\vrule
&\ch\hfil
&\multispan 6\hfil$\mathstrut\hskip5pt\nabla^Z$\hfil\vrule&\hfil\hbox{Divisor}\cr
\noalign{\hrule\vskip3pt\hrule}
v_1&0&&-1&0&2&3&&K_+\cr
v_2&0&&0&-1&2&3&&K_-\cr
v_3&-24&&0&0&-1&0&&2H^Z\cr
v_4&-57&&0&0&0&-1&&3H^Z\cr
{1\over 2}(v_1{+}v_6)={1\over 2}(v_2{+}v_5)&1&&0&0&2&3&&B^Z\cr
v_5&0&&0&1&2&3&&K_-\cr
v_6&0&&1&2&2&3&&K_+\cr
\noalign{\hrule\vskip3pt\hrule}
\multispan{9}\vrule height15pt depth 10pt \hfil $(h^{11},h^{21})~=~(3,243)~,
~~\ch_E~=~-480~,~~(\d,\tilde\d)~=~(0,0)$\hfil\vrule\cr
\multispan{9}\vrule height10pt depth7pt\hfil $H^Z=B^Z+2K_+ + 2K_-$\hfil\vrule\cr
\noalign{\hrule}
}}}
$$\divisorsZ$$
\vskip10pt
The manifold $Z$ is elliptically fibered over a base $B^Z=\cp1\times\cp1$. We denote
these two $\cp1's$ by $\ca{L}_+$ and $\ca{L}_-$ and we see that the elliptic fibers over
$\ca{L}_\pm$ form K3-surfaces 
 $$
K_\pm ~=~(\ca{E},\ca{L}_\pm)~.$$
The fan for $Z$ is 
 $$
\S^Z = (v_1+v_6)(v_2+v_5)\S^{\ca{E}}~,~~\hbox{with}~~\S^{\ca{E}}= B^Z(v_3+v_4) + v_3v_4~.$$
Given the fan SCHUBERT immediately provides the intersection numbers:
 $$
(B^Z)^3~=~8~,~~~~(B^Z)^2.K_\pm~=~-2~,~~~~B^ZK_+K_-~=~1,~~~~K_\pm^2~=~0~.$$

Finding the \K\ and Mori cones of $Z$ is an easy exercise in the method of piecewise
linear functions (though more elementary procedures work in this case also).
In a notation analogous to that introduced in Sect. 5.2 one is led to the following inequalities
 $$\eqalign{
0\leq& -2m_b+m_1+m_6\cr
0\leq& -2m_b+m_2+m_5\cr
0\leq& \hphantom{-2}m_b-6m_0+2m_3+3m_4\cr
0\leq&-12m_0+m_1+4m_3+6m_4+m_6\cr
0\leq&-12m_0+m_2+4m_3+6m_4+m_5~.\cr}$$
A basis is provided by the first three inequalities so taking a basis of divisors to be
$(B^Z, v_6,v_5)=(B^Z,K_+,K_-)$ we see that the curves that generate the Mori cone are
 $$\eqalign{
&(-2,\- 1,\- 0)\cr
&(-2,\- 0,\- 1)\cr
&(\- 1,\- 0,\- 0)\cr}$$
which are the curves
 $$
\ca{L}_+~=~B^ZK_+~,~~~~\ca{L}_-~=~B^ZK_-~,~~~~\ca{E}~=~K_+K_-~.$$
The generators of the \K\ cone are the divisors $(K_-,K_+,H^Z)$ that are dual to these curves.

We may write the \K-form as a linear combination of the generators
$$
J^Z~=~tH^Z + u_{-}K_{+} + u_{+}K_{-} ~.$$
Written this way the parameters are the volumes of the dual curves:
 $$
J^Z\ca{E}~=~t~,~~~~J^Z\ca{L}_\pm~=~u_\pm~.$$
We record here also the volume of $Z$ itself as well as that of the base $B^Z$ and that
of the~$K_\pm$.
 $$\displaylines{
{1\over 3!}(J^Z)^3 = t\left[ u_{+}u_{-} + t(u_{+} + u_{-}) + {4\over 3}t^2 \right]\cropen{5pt}
{1\over 2!}(J^Z)^2 B^Z = u_{+}u_{-}~,~~~~{1\over 2!}(J^Z)^2 K_\pm = t(u_\pm + t)~.\cr} $$
Note that, owing to the fact that the elliptic fiber varies over the base, the volume of $Z$
is not simply the volume of the base multiplied by the volume  of the fiber unless
$t$ is small.
\midinsert
\def\Zfig{\vbox{\vskip0pt\hbox{\hskip0pt\epsfxsize=3.5truein\epsfbox{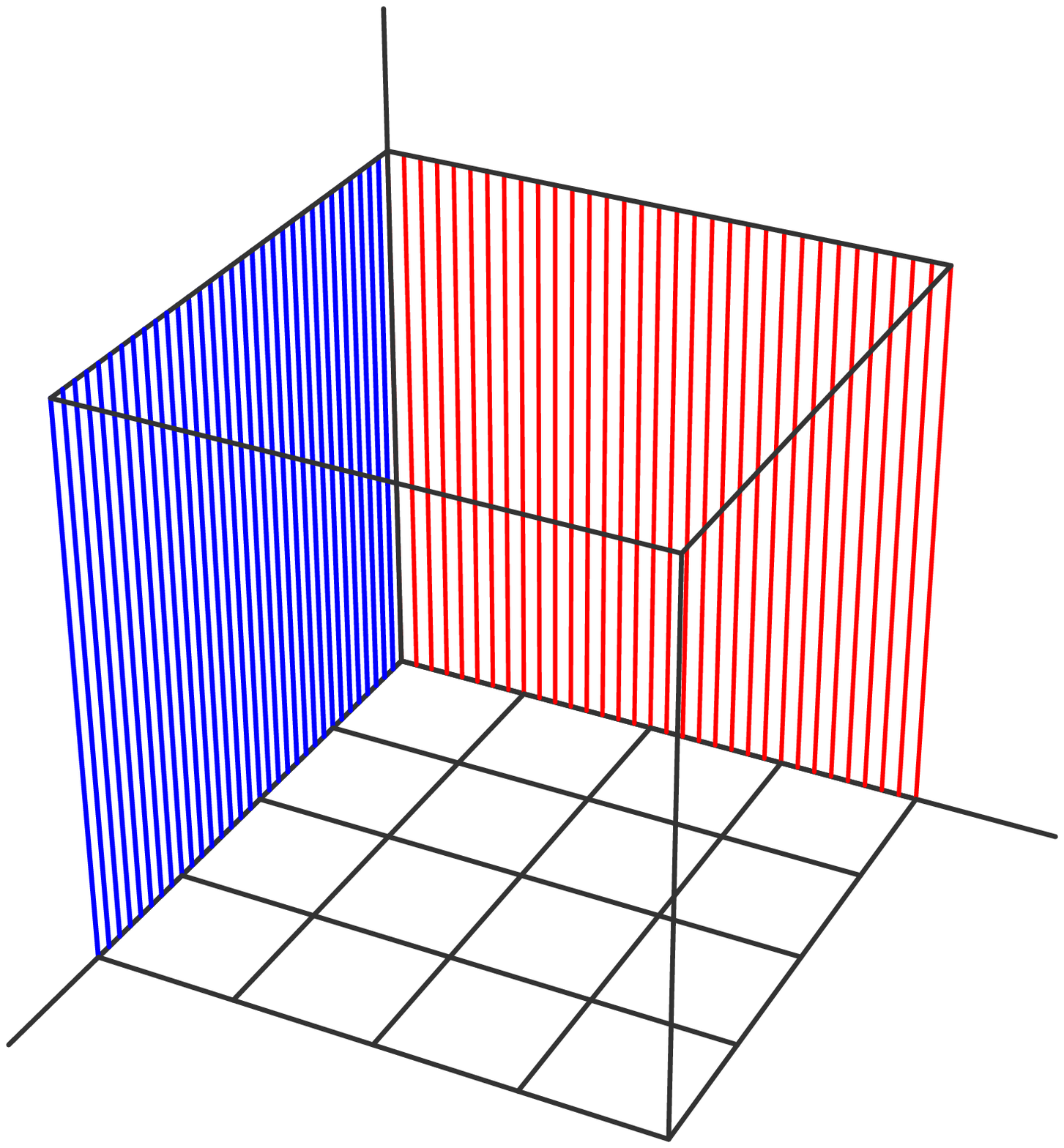}}}}
\figbox{\Zfig\vskip-20pt}{\figlabel{Zfig}}{A sketch of $Z$ as an elliptic fibration
over $\ca{L}_+{\times}\ca{L}_-$ and as a $K3$-fibration in two ways as $(K_+,\ca{L}_-)$
and $(K_-,\ca{L}_+)$.}
\place{3}{4.3}{$\ca{E}$}
\place{1.4}{1.5}{$\ca{L}_+$}
\place{4.8}{2.2}{$\ca{L}_-$}
\place{2.1}{3.8}{$K_+$}
\place{3.7}{3.9}{$K_-$}
\endinsert
\subsection{Projection to $B^Z$}
As discussed in Section~\chapref{heterotic} the projection to $Z$ corresponds to
projecting out the third component of the points of $\nabla^X$. In reality the
projection to $Z$ exists in only a limited sense. There is a well defined projection
to $B^Z$ which is a section of the fibration $B=(\cp1, B^Z)$. The section is not
unique nevertheless there are projections onto each of these sections. 

We may also
project the divisors of $X$ onto the divisors of $Z$. This proceeds in the following
way. The projection of $C_2\simeq (0,0,-1,1,2)$ is $(0,0,1,2)$ which is interior to
a codimension one face of $\nabla^Z$. We therefore take $C_2$ to project to zero. In
an analogous way we see that we should also take $E_3$ to project to zero.
Now we see from the polyhedra that we should take
 $$
H\to H^Z~,~~\hbox{and}~~Y^{\pm}\to K_{\pm}~. $$
Now $H= B+C_1+C_2+E_1+E_2+2Y^{+}+2Y^{-}$ and $H^Z=B^Z+2K_{+}+2K_{-}$. So we take also
 $$
B\to B^Z~~\hbox{and}~~C_i\to 0~,~~E_j\to 0~~\hbox{for all}~i,j.$$
Now there is an element of choice in what we wish to call the preimage of $B^Z$
under this projection since we are free to add multiples of the divisors that
project to zero. For the intersection calculation that follows it is sufficient to
take this preimage to be $\hat{B} = B + C_1 + E_1$. We check that we obtain the
correct values for the intersection numbers on $Z$:
 $$\vbox{\offinterlineskip\tabskip=0pt\halign{%
$#$\hfil&$#$\hfil&$#$\hfil&$#$\hfil&#\cr
(B^Z)^3 &= \hat{B}^2 B^Z &= \hat{B}^2 BE_1 &=\- 8\cropen{3pt}
(B^Z)^2 K_{\pm} &= \hat{B} Y^{\pm} B^Z &= \hat{B} Y^{\pm} BE_1 &= -2\cropen{5pt}
B^Z K_{+} K_{-} &= Y^{+} Y^{-} B^Z &= Y^{+} Y^{-} BE_1 &=\- 1~.\cr
}} $$
Note that we could take instead $B^Z=BC_1$ and these intersection numbers would
still be correct. Now observe that since the images of $F$ and $G$ under the
projection are both multiples of $K_-$ we should set
 $$
F\to \a K_-~~\hbox{and}~~G\to (1-\a) K_-$$
for some $\a$. It turns out that $\a$ is 0 or 1 depending on whether we take $B^Z$
to be $BE_1$ or $BC_1$ since in the first case $B^Z$ intersects $G$ but not $F$
while in the second $B^Z$ intersects $F$ but not $G$. 
\newpage
\section{appY}{The Divisors for the Spaces 
$\hbox{\fourteenmath Y\hskip3pt}^\pm$} 
We record here Tables for the divisors of $Y^+$ and
$Y^-$. It is evident from the topological numbers that the two manifolds $Y^\pm$ are
different.
\vskip5pt
\def\divisorsYplus{%
\vbox{\def\skip{\hskip5pt}\def\space{\hphantom{1}}\def\nostar{\hphantom{^*}}
\offinterlineskip\tabskip=0pt\halign{%
\vrule height12pt depth8pt\skip$##$\skip\hfil\vrule&\hfil\skip$##$\skip\vrule&
\hfil\skip(##&\hfil##,&\hfil##,&\hfil##,&\hfil##&##)\skip\hfil\vrule&
\skip$##$\skip\hfil\vrule\cr
\noalign{\hrule}
\omit\vrule height18pt depth14pt\hfil\titlebox{Relation}{to vertices}\hfil\vrule
&\chi\hfil
&\multispan 6\hfil$\mathstrut\hskip5pt\nabla^{Y^+}$\hfil\vrule
&\hfil\hbox{Divisor}\cr
\noalign{\hrule\vskip3pt\hrule}
V^{+}_2&0&&-1&0&2&3&&Y^{+} = F^{+} + G^{+}\cr
V^{+}_3&1&&0&-1&2&3&&C^{+}_1\cr
V^{+}_9&1&&0&-1&1&2&&C^{+}_2\cr
V^{+}_4&-13&&0&0&-1&0&&2H^{+} + E^{+}_3 - C^{+}_2\cr  
V^{+}_5&-38&&0&0&0&-1&&3H^{+} + E^{+}_3 - C^{+}_2\cr 
{1\over 2}(V^{+}_3 {+} E^{+}_1)&1&&0&0&2&3&&B^{+}\cr
{1\over 2}(V^{+}_2 {+} V_F)&1&&0&1&2&3&&E^{+}_1\cr
V^{+}_8&1&&0&2&2&3&&E^{+}_2\cr
{1\over 2}(V^{+}_5{+}V^{+}_8)&2&&0&1&1&1&&E^{+}_3=C^{+}_1+C^{+}_2-(E^{+}_1+2E^{+}_2+2F^{+}+3G^{+})\cr
V_F&1&&1&2&2&3&&F^{+}\cr
V^{+}_7&1&&1&3&2&3&&G^{+}\cr
\noalign{\hrule\vskip3pt\hrule}
\multispan 9\vrule height15pt depth10pt
\hfil$ H^{+}~=~B^{+}+C^{+}_1+C^{+}_2+E^{+}_1+E^{+}_2+2Y^{+}$~ \hfil\vrule\cr
\multispan 9\vrule height15pt depth10pt
\hfil$(h^{11},h^{21})~=~(7,169)~,$ \hfil $(\d,\tilde\d)~=~(0,1)~,$\hfil 
$\ch_E ~=~ -324$\hfil\vrule\cr
\noalign{\hrule}
}}}
$$\divisorsYplus$$
\centerline{Table B1: The divisors for $Y^+$.}
\newpage
\def\divisorsYminus{%
\vbox{\def\skip{\hskip5pt}\def\space{\hphantom{1}}\def\nostar{\hphantom{^*}}
\offinterlineskip\tabskip=0pt\halign{%
\vrule height12pt depth8pt\skip$##$\skip\hfil\vrule&\hfil\skip$##$\skip\vrule&
\hfil\skip(##&\hfil##,&\hfil##,&\hfil##,&\hfil##&##)\skip\hfil\vrule&
\skip$##$\skip\hfil\vrule\cr
\noalign{\hrule}
\omit\vrule height18pt depth14pt\hfil\titlebox{Relation}{to vertices}\hfil\vrule
&\chi\hfil&\multispan 6\hfil$\mathstrut\hskip5pt\nabla^{Y^-}$\hfil\vrule
&\hfil\hbox{Divisor}\cr
\noalign{\hrule\vskip3pt\hrule}
V^{-}_1&0&&-1&0&2&3&&Y^-\cr
V^{-}_3&1&&0&-1&2&3&&C^{-}_1\cr
V^{-}_9&1&&0&-1&1&2&&C^{-}_2\cr
V^{-}_4&-14&&0&0&-1&0&&2H^{-} + E^{-}_3 - C^{-}_2\cr  
V^{-}_5&-45&&0&0&0&-1&&3H^{-} + E^{-}_3 - C^{-}_2\cr 
{1\over 3}(2V^{-}_3 {+} E^{-}_2)&1&&0&0&2&3&&B^{-}\cr
{1\over 2}(B^{-} {+} E^{-}_2)&1&&0&1&2&3&&E^{-}_1\cr
{1\over 2}(V^{-}_1 {+} V^{-}_6)&1&&0&2&2&3&&E^{-}_2\cr
{1\over 2}(V^{-}_5 {+} E^{-}_2)&-1&&0&1&1&1&&E^{-}_3=C^{-}_1+C^{-}_2-(E^{-}_1+2E^{-}_2+4Y^{-})\cr
V^{-}_6&0&&1&4&2&3&&Y^-\cr
\noalign{\hrule\vskip3pt\hrule}
\multispan 9\vrule height15pt depth10pt
\hfil$ H^{-}~=~B^{-}+C^{-}_1+C^{-}_2+E^{-}_1+E^{-}_2+2Y^{-}$~ \hfil\vrule\cr
\multispan 9\vrule height15pt depth10pt
\hfil$(h^{11},h^{21})~=~(8,194)~,$ \hfil $(\d,\tilde\d)~=~(0,2)~,$\hfil 
$\ch_E ~=~ -372$\hfil\vrule\cr
\noalign{\hrule}
}}}
$$\divisorsYminus$$
\centerline{Table B2: The divisors for $Y^-$.}
\newpage
\frenchspacing
\immediate\closeout\referencewrite\referenceopenfalse
\line{\fourteenbold\hfil References\hfil}\bigskip\parindent=0pt\input referenc.texauxil
\bye